\documentclass[pre,amsmath]{revtex4-2}
\usepackage[pdftex]{graphicx} 
\usepackage{color}
\usepackage{bm}
\usepackage{epsfig}
\usepackage{latexsym}
\usepackage{nameref,hyperref}
\begin{document}
\newcommand{\be}{\begin{equation}}
\newcommand{\ee}{\end{equation}}
\newcommand{\bea}{\begin{eqnarray}}
\newcommand{\eea}{\end{eqnarray}}
\newcommand{\RN}[1]{\textup{\uppercase\expandafter{\romannumeral#1}}}
\title{Bacterial Chemotaxis in a Traveling Wave Attractant Environment }
\author{Shobhan Dev Mandal and Sakuntala Chatterjee}
\affiliation{Department of Physics of Complex Systems, S. N. Bose National Centre for Basic Sciences, Block JD, Sector 3, Salt Lake, Kolkata 700106, India.}
\begin{abstract}
We study single cell {\sl E.coli} chemotaxis in a spatio-temporally varying attractant environment. Modeling the attractant concentration in the form of a traveling sine wave, we measure in our simulations, the chemotactic drift velocity of the cell for different propagation speed of the attractant wave. We find a highly non-trivial dependence where the chemotactic drift velocity changes sign, and also shows multiple peaks. For slowly moving attractant wave, drift velocity is negative, {\sl i.e.} the drift motion is directed opposite to wave propagation. As the wave speed increases, drift velocity shows a negative peak, then changes sign, reaches a positive peak and finally becomes zero when the wave moves too fast for the cell to respond. We explain this rich behavior from the difference in attractant gradient perceived by the cell during its run along the propagation direction and opposite to it. In particular, when the cell moves in the same direction as the wave, the relative velocity of the cell with respect to the wave becomes zero when the wave speed matches the run speed. In this limit, the cell is able to ride the wave and experiences no concentration gradient during these runs. On the contrary, for runs in the opposite direction, no such effect is present and the effective gradient increases monotonically with the wave speed. We show, using detailed quantitative measurements, how this difference gives rise to the counter-intuitive behavior of chemotactic drift velocity described above.
\end{abstract}
\maketitle

\section{Introduction}

                                                                                                                                                                                                            Chemotaxis is the directed motion of an organism in response to a chemical signal coming from its environment \cite{eisenbach}. In many biological systems, chemotaxis takes place in a chemical environment which itself shows spatio-temporal variation. Examples include quorum sensing by bacterial cells \cite{daniels2004quorum}, aggregate formation by Dictyostelium discoideum in nutrient deficient conditions \cite{dubravcic2014evolutionarily, cai2012analysis, goldstein1996traveling}, or recruitment of neutrophil cells to the inflammation sites in mammals \cite{afonso2012ltb4}. In certain situations, due to interaction between the cells and the attractant environment, moving wavefronts are generated \cite{ishida2025traveling}. In particular, attractant degradation can give rise to local gradient and as the cells perform chemotaxis to reach regions of high attractant levels, this gradient is enhanced which further induces chemotaxis \cite{tweedy2016self}. This positive feedback gives rise to a moving wavefront. Apart from degradation, sometimes attractant concentration is also self-amplified and this information is relayed between the cells. This mechanism can generate pulsatile waves     \cite{goldbeter1997modelling}. Motivated by these, in this paper we study chemotaxis in the presence of spatio-temporally varying attractant environment. Specifically, we focus on {\sl E.coli} chemotaxis, which has received a lot of research attention in physics as well as in biology \cite{2008cbergoli, tu2013quantitative}. The transmembrane chemoreceptors in an {\sl E.coli} cell change their conformation or `activity' state upon binding with attractant molecules. The resulting activity modulates the run-tumble motion of the cell causing it to move in the direction of increasing attractant level in the medium. A well-characterized signaling pathway makes {\sl E.coli} a model system for studying many chemotaxis-related phenomena \cite{2008cbergoli}.

There have been a number of useful theoretical and experimental studies on {\sl E.coli} chemotaxis in dynamical attractant environment in the recent past \cite{tu2008modeling, shimizu2010modular, lazova2011response, zhu2012frequency, jiang2010quantitative, li2017barrier}. In one of the pioneering studies in this direction \cite{tu2008modeling}, a nutrient concentration profile varying exponentially in time was considered and the dependence of receptor activity on the exponential ramp rate was measured both experimentally and theoretically. In a subsequent study \cite{shimizu2010modular} the receptor activity was measured in response to an attractant signal that oscillated with time. It was found that receptor activity also shows similar oscillations, but with a phase delay \cite{shimizu2010modular, lazova2011response}. Similar phase delay was also reported for temporal variation of cell density in an  oscillating attractant environment \cite{zhu2012frequency}. While the cell density varies in sync with the attractant level for small oscillation frequency, they become out of phase for fast oscillations \cite{zhu2012frequency}. Damped oscillation in cell density was observed in presence of an attractant profile whose spatial gradient oscillates with time \cite{jiang2010quantitative}, when the frequency of oscillation exceeds a threshold value. In a particularly interesting development \cite{li2017barrier}, a traveling wave attractant profile was considered in a microfluidic channel and population averaged drift velocity was measured. It was shown that for a slowly moving attractant wave, drift velocity increases with wave speed but as the wave moves faster, drift velocity shows a sharp fall. This experimental observation was explained using Kramers reaction rate theory within a coarse-grained model \cite{li2017barrier}.

In the presence of a time-periodic attractant environment, most chemotactic response functions like cell density, receptor activity, etc. also show   time-periodic variation with the same periodicity as the attractant wave. In most of the earlier studies, such response has been measured as a function of time, over a complete time-period of attractant oscillation. However, important insights about the system may be found from time-averaged measurements as well, where averaging is done over all times, irrespective of the phase of the periodic cycle. In this work, we perform such measurements. We consider single cell E.coli chemotaxis with attractant concentration profile in the form of a propagating sinusoidal wave, $[L](x,t)=[L]_0+A\sin \dfrac{2\pi}{\lambda}(v_wt-x)$ with wavelength $\lambda$ and propagation speed $v_w$. Performing numerical simulations on a quantitative model of intracellular signaling network, we measure time-averaged drift velocity of the cell in steady state and focus on its variation with $v_w$. For $v_w=0$ the attractant profile represents a static sine wave for which the cell tends to accumulate near the concentration maxima in the long time limit. No net chemotactic drift is expected in this case. Similarly, for very large $v_w$ the attractant wave moves too fast for the cell to respond. In this case, the cell experiences an average attractant concentration which is homogeneous everywhere. Drift velocity vanishes in this limit too. For intermediate values of $v_w$ we find a highly non-trivial behavior of drift velocity. More specifically, depending on whether the propagation speed of the attractant wave is smaller or larger than the run speed $v$ of the cell, the time-averaged chemotactic drift velocity shows qualitatively different behavior. Our simulations show that for  $v_w \ll v $ drift velocity is negative, {\sl i.e.} directed in the opposite direction of wave propagation. As $v_w$ increases, drift velocity reaches a negative minimum. For $v_w \gtrsim v $ drift velocity changes sign, and becomes positive. It reaches a positive maximum with $v_w$ in this range and finally becomes zero as $v_w$ becomes significantly larger than $v$.

\begin{figure}
\includegraphics[scale=1.5]{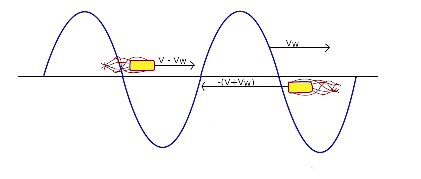}
\caption{Schematic diagram of an {\sl E.coli} cell moving in a traveling wave attractant environment. Left-movers and right-movers have different relative velocities with respect to the traveling wave.}\label{fig:mov}
\end{figure}

To explain this interesting behavior, we focus on the different attractant environments experienced by the cell when it is moving in the same direction, or in the opposite direction of wave propagation. Without any loss of generality we consider here rightward propagation of the  attractant wave. This means during a rightward run of the cell, the relative velocity of the cell with respect to the moving wave is $(v-v_w)$ (see Fig. \ref{fig:mov}). Starting from small values, as $v_w$ increases, the cell effectively experiences a slower attractant wave, or equivalently, a smaller attractant gradient during a rightward run. For $v_w=v$ the gradient becomes zero. At this special point, the cell `rides' the wave and sees no change in attractant level as long as it is running rightward. If a new rightward run started when the cell was in a low concentration zone of the attractant wave, throughout the run it continues to sense the same low concentration. For $v_w > v$ the attractant wave overtakes the right-moving cell, which therefore effectively feels an attractant wave moving backward (leftward). However, for a left-moving cell the scenario is different. In this case, the relative velocity of the cell during the leftward run is $-(v+v_w)$ and as $v_w$ increases, the cell experiences a steeper attractant gradient for all ranges of values of $v_w$. Chemotactic drift arises due to difference in rightward and leftward run durations. The different attractant environments experienced during leftward and rightward runs, for different values of $v_w$, has important consequences on these run durations. We show that this difference is finally responsible for the observed variation of chemotactic drift velocity. 

In the next section we describe the model in details. Our results for moderate and large values of $\lambda$ appear in Secs. \ref{sec:lambda} and \ref{sec:large}. In Sec. \ref{sec:con} we summarize our results and make concluding remarks.


\section{Model Description} \label{sec:model}

\begin{figure}
\includegraphics[scale=1.3]{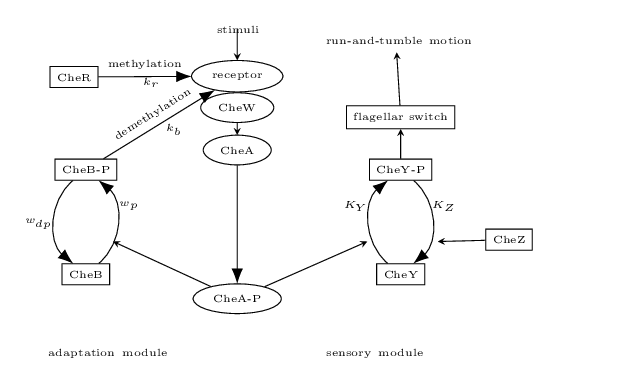}
\caption{A schematic diagram of chemotactic signaling network inside an {\sl E.coli} cell.} \label{fig:path}
\end{figure}

 In Fig. \ref{fig:path} we show a schematic diagram of the signaling network inside the cell. This network has two principal modules: sensing and adaptation. While the sensing module is responsible for modulating the run-tumble motion according to the local attractant concentration in the extracellular environment, the adaptation module controls the methylation levels of the receptors to ensure that the receptor activity does not deviate too much from its adapted value. In the active state, the receptors cause auto-phosphorylation of cytoplasmic protein CheA, which then donates its phosphate group to two other proteins, CheB and CheY. In the phosphorylated form CheY-P induces tumbling of the cell, and CheB-P causes demethylation of the receptors. When receptors bind to attractant molecules, their activity decreases, which in turn decreases the tumbling rate, and the cell swims smoothly. Inactive receptors, on the other hand, get methylated which raises the activity. When attractant molecules are scarce, the activity of the receptors remains high. This causes high tumbling rate and short runs of the cell. Active receptors again get demethylated which lowers their activity.  

We use a stochastic theoretical model to describe this signaling pathway \cite{shobhan,mandal2022effectre,mandal2022effect}. Our model consists of three main parts: (A) switching of chemoreceptor activity, (B) methylation and demethylation of chemoreceptors by enzymes and (C) run-tumble motility of the cell. Below we describe each part in detail. 

\subsection*{Switching of chemoreceptor activity }

(A) In an {\sl E.coli} cell there are few thousand transmembrane receptors \cite{pontius2013adaptation,briegel2012bacterial}. Two receptor monomers form a receptor dimer, which can be in two different states: active and inactive \cite{liu2012molecular} and the switching between these two states happen stochastically. We use Monod-Wyman-Changeux (MWC) model \cite{mello2005allosteric, monod1965nature, keymer2006chemosensing} to describe the switching dynamics between the two activity states.
According to MWC model, the switching rate between active and inactive states is governed by their free energy difference $\epsilon$, which has the form (in the unit of $K_BT$)
\begin{equation}
\epsilon([L],m)=log\frac{1+[L]/K_{min}}{1+[L]/K_{max}}+\epsilon_0-\epsilon_1m
\end{equation}
where  $[L]$ is local ligand concentration in the cell environment which may change with space and time. $m$ is the methylation level of the dimer. $m$ can take any integer value between $0$ and $8$ \cite{dufour2014limits, pontius2013adaptation, frankel2014adaptability, long2017feedback}. Here, $K_{min}$ and $K_{max}$ are ligand dissociation constant for inactive and active receptors, respectively. They set the range of sensitivity: the cell can sense any concentration in the range $K_{min} < [L] < K_{max}$. The values of the parameters $\epsilon_0$ and $\epsilon_1$ are listed in Table \ref{table}. 

Due to cooperative interaction among the receptors, three dimers come together to form a trimer of dimers and $n$ such trimers form a receptor cluster or signaling team  \cite{briegel2012bacterial}. All dimers in a signaling team switch their activity in unison, which results in significant signal amplification. According to MWC model, the total free energy of a receptor cluster containing $3n$ dimers is
\begin{equation}
F=3n \log\frac{1+[L]/K_{min}}{1+[L]/K_{max}}+3n\epsilon_0-\sum_{i=1}^{3n}\epsilon_1 m_i  \label{eq:nrg}
\end{equation}
where $m_i$ is the methylation level of the $i$-th dimer in the cluster. $F$ controls the transition between active and inactive states of the cluster. In our model, using the principle of local detailed balance, the rate of transition from active to inactive state is taken to be $w_a/[1+\exp(-F)]$ and the reverse transition rate is $w_a/[1+\exp(F)]$, where $w_a$ is the switching time-scale whose value is given in Table \ref{table} \cite{mandal2022effectre, colin2017multiple, chatterjee2022short}.  For the sake of simplicity, we assume here all receptor clusters are of the same size.

\subsection*{Methylation and demethylation of chemoreceptors by enzymes }

(B) Methylation dynamics of receptors are controlled by two enzymes present in the cell, CheR and CheB, which are known as methylating and demethylating enzymes, respectively. One enzyme molecule can bind to one dimer at a time. We present the details of binding-unbinding kinetics of these enzymes in Appendix \ref{app:model}. In a bound state, a CheR molecule can increase methylation level $m$ of a receptor dimer by one unit if and only if the receptor dimer is inactive and with $m <8$. The demethylating enzyme CheB first undergoes phosphorylation (as shown in Fig. \ref{fig:path}) in presence of an active receptor cluster. The phosphorylated form CheB-P can bind to a receptor dimer and once bound, it can lower its methylation level $m$ by one unit provided the receptor dimer is active and $m >0$. From the expression for free energy in Eq. \ref{eq:nrg} it follows that as $m$ increases, activity also increases. Since inactive receptors get methylated and active receptors get demethylated, the methylation-demethylation reactions therefore act as a negative feedback system which prevents large deviation of activity from its adapted value.

\subsection*{Run-tumble motility of the cell}

\begin{figure}
\includegraphics[scale=0.7]{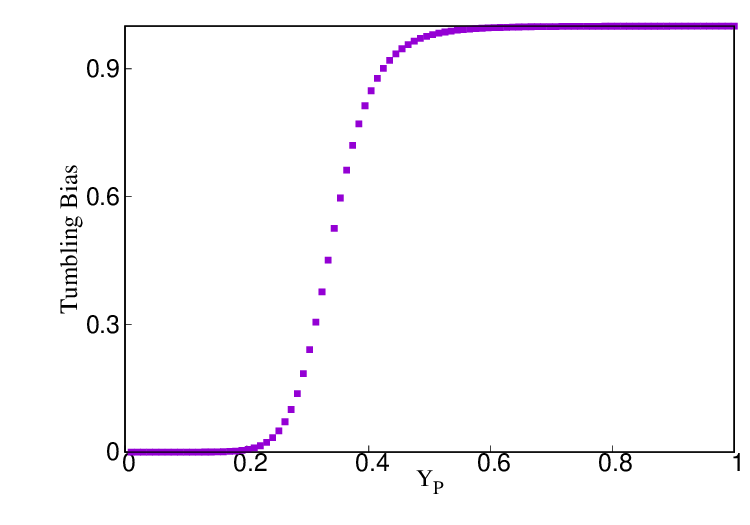}
\caption{Tumbling bias as a function of normalized CheY-P concentration $Y_P$. Tumbling bias shows a step-like dependence on $Y_P$.}
\label{fig:cwb}
\end{figure}

(C) As shown in Fig. \ref{fig:path}, active receptors cause phosphorylation of CheY. In the phosphorylated form CheY-P binds to the flagellar motors and causes the motors to rotate in the clockwise direction. This induces tumbling of the cell. The phosphorylation-dephosphorylation dynamics of CheY follows the equation: $\dfrac{dY_P}{dt} = K_Y a (1-Y_P) -K_Z Y_P $, where $Y_P$ denotes the ratio of concentration of CheY-P and CheY, and $a$ denotes the fraction of receptors in the active state, and $K_Y$ and $K_Z$ are rate constants for phosphorylation and dephosphorylation reactions whose values are listed in Table \ref{table}. We do not explicitly model this dynamics but assume that (de)phosphorylation process of CheY is quite fast and therefore the phosphorylated fraction  of CheY proteins instantly reaches its steady state value for a given receptor activity level. In other words, $Y_P = \dfrac{a}{a+K_z/K_Y}$  \cite{flores2012signaling, jiang2010quantitative, dev2018optimal}. However, we have verified (data not shown here) that even when we relax this assumption, and calculate $Y_P(t)$ after numerically integrating the above equation, our conclusions are not affected. The cell stochastically switches between its run state and tumble state, and the switching rate depends on $Y_P$. In our model, a running cell can tumble with a rate $\omega\exp(-G)$, where $G=\Delta_1-\frac{\Delta_2}{1+Y_0/Y_P}$, and $\omega, Y_0$ are constants whose values are given in Table \ref{table} \cite{sneddon2012stochastic, dufour2014limits, micali2017drift}. The reverse transition happens with rate $\omega\exp(G)$. In Fig. \ref{fig:cwb} we have shown the variation of tumbling bias as a function of CheY-P concentration. 

Using the above model, we perform stochastic simulations using Monte Carlo method to study chemotaxis in one and two spatial dimensions. We use time-step $dt=0.01s$ in our simulations. Any reaction or switching process, described above, with a specific rate $w$ can take place with probability $w dt$ in the time-step. For those cases, where the rate is larger than $1/dt$, we assign a probability $1$. In our model, with the present choice of $dt$, only the rebinding rate $w_{rb}$ is larger than $1/dt$ \cite{pontius2013adaptation}. After each tumble the cell chooses a direction at random and runs in that direction. In a one dimensional run-and-tumble motion, the cell can move only leftward or rightward with a fixed run speed $v$. This means in a single time-step $dt$ the cell covers $\pm vdt$ distance. Therefore, in one dimension, the cell moves on a lattice with lattice spacing $vdt$. But in two dimensions there is an angle $\theta$ between two consecutive runs that can take any value between $0$ and $2 \pi$ with uniform probability. In this case, we also include a rotational diffusion, where $\theta$ changes slowly during a run, which gives rise to a gradual bending of the trajectory, as seen in experiments \cite{berg1972chemotaxis}.

\subsection*{Traveling wave attractant profile}
   
We have considered an attractant profile which varies in space and time. We are mainly interested in the long time behavior of the cell and consider a traveling wave attractant concentration profile 
\be 
[L](x,t)=[L]_0+A\sin \frac{2\pi}{\lambda}(v_wt-x). \label{eq:cxt}
\ee
Without any loss of generality, we consider $v_w > 0$ here. For $v_w=0$ we have a static attractant profile with sinusoidal spatial variation. The cell will tend to localize at the maxima of the sine wave. The chemotactic drift velocity is expected to be zero in this limit. However, for $v_w \neq 0$ the position of maxima and minima change with time and our simulations show that the chemotactic response of the cell depends strongly on the wavelength $\lambda$. We consider $\lambda$ values much larger than the average distance $v \tau$ travelled by the cell during a run. For our choice of $\lambda = 200 \mu m$, the ratio $\lambda / v \tau \sim 4.3$. We describe this as a moderate $\lambda$ and the case $\lambda = 2000 \mu m$ is described as large $\lambda$. Various different quantities which can be measured to characterize the cell response, show different qualitative behavior for moderate and large $\lambda$. Below we present our results for these two cases separately.  For most parts of the paper, we consider run-tumble motion in one spatial dimension, in which case the cell moves in a periodic ring of length $L$. In Appendix B we consider two dimensional motion of the cell in a region of size $L_x \times L_y$ with periodic boundary conditions assumed in both $x$ and $y$ directions.

All our measurements are performed in the long time limit when the system has reached a (time-periodic) steady state. Starting from random initial values of the dynamical variables in our model, like number of active receptors, methylation level of each dimer, position of the cell, etc., we perform stochastic time-evolution and update these variables with time. After a large enough number of updates, the system reaches steady state when the time-averaged quantities like mean run duration, drift velocity {\sl etc.} do not increase or decrease with time anymore, but merely fluctuate around some steady value, and the time-periodic quantities like cell density $P(x,t)$, activity $a(x,t)$, {\sl etc.} show traveling wave form with the same periodicity as $[L](x,t)$. In our simulations we observe that the cell reaches this state within $\sim 200$ run-tumble events. Once in steady state, we start our measurements, which we average over at least $10^6$ number of steady state configurations.

\section{Chemotactic response for moderate wavelength} \label{sec:lambda}

In this section, we present our results for $\lambda = 200 \mu m$. We measure chemotactic drift velocity $V$ for different values of the propagation speed $v_w$. For the present run-tumble motion, we define the drift velocity as the net displacement of the cell during a run, divided by average run duration \cite{de2004chemotaxis, Locsei2007, Taktikos2013, Thornton2020, chatterjee2011chemotaxis, dev2018optimal, shobhan}. An alternative definition of drift velocity as mean displacement per unit time can also be used and shows similar behavior (see data in Fig. \ref{fig:u}). Here, we present a detailed protocol of our measurement. After reaching steady state, we observe the system for a long enough time window ${\mathcal T} $. Let $N_R(x)$ be the total number of rightward runs starting from position $x$ during our observation window. Let $d_R(x)$ be the total duration of these runs. Similarly, $N_L(x)$ and $d_L(x)$ can be defined for leftward runs. Then the mean duration $\tau$ of all runs measured over the observation window ${\mathcal T} $ is
\begin{equation}
\tau = \frac{\int dx [d_R(x) +d_L(x)]}{\int dx [N_R(x)+N_L(x)]},  \label{eq:tau}
\end{equation}
where the numerator is the total duration of all runs (starting from any position) and the denominator is the total number of the runs. The net displacement $\Delta$ of the cell in this observation window has the form
\begin{equation}
\Delta = \frac{v \int dx [d_R(x)-d_L(x)]}{\int dx [N_R(x)+N_L(x)]} 
\label{eq:del}
\end{equation}
where the numerator gives the difference between total rightward displacement and total leftward displacement. The denominator is the total number of runs, as before. The drift velocity $V$ is then given by 
\begin{equation}
V = \frac{\Delta}{\tau} = \frac{v \int dx [d_R(x)-d_L(x)] }{\int dx [d_R(x) +d_L(x)] } \label{eq.vdef}
\end{equation}

\begin{figure}
\includegraphics[scale=0.7]{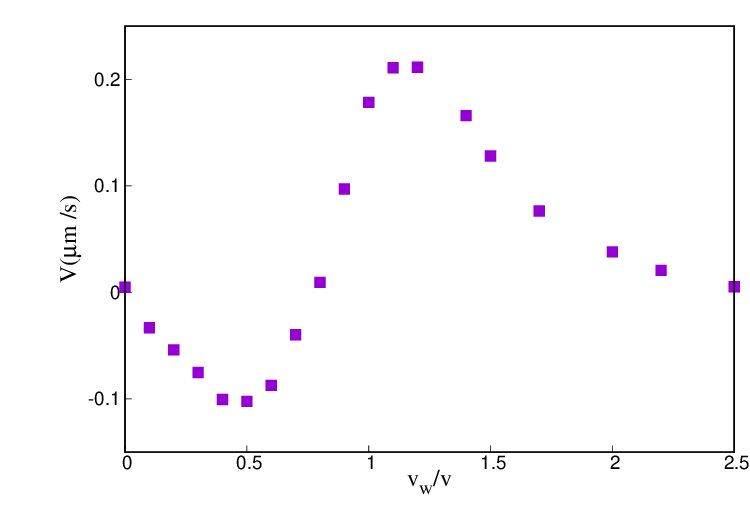}
\caption{Variation of chemotactic drift velocity $V$ with propagation speed $v_w$ of the attractant wave. $v_w$ has been scaled by the constant run speed $v$ of the cell. $V$ vanishes for very small and very large $v_w$. For intermediate $v_w$ we find $V$ shows a negative minimum, crosses zero, reaches a positive maximum and finally falls to zero.  These data are for one spatial dimension. The maximum error-bar in our measurement is $0.001 \mu m /s$. Here, we have used wavelength $\lambda = 200 \mu m$, and oscillation amplitude $A =5 \mu M$. All other simulation parameters are listed in Table \ref{table}. } \label{fig:v100}
\end{figure}

We find rich behavior of $V$ as $v_w$ is varied. We present our data in Fig. \ref{fig:v100}. As expected, $V$ vanishes for $v_w=0$ which corresponds to a static wave. For very large $v_w$ when the wave moves too fast, the cell is unable to sense the spatio-temporal variation and $V$ vanishes in this limit too. For finite $v_w$, however, very interesting variation in $V$ is observed. Starting from zero, as $v_w$ increases, $V$ first becomes negative, reaches a minimum, then increases again, becomes positive, reaches a peak and finally vanishes for very large $v_w$. Therefore, by tuning $v_w$ one can have the net drift of the cell in the same direction, or even in the opposite direction of attractant wave propagation. Regarding the magnitude of drift velocity, it may be useful to compare with the typical drift velocity observed in a related system. It follows from Eq. \ref{eq:cxt} that the maximum value of the concentration gradient faced by the cell is $2 \pi A / \lambda$. We consider a static attractant environment with a constant spatial gradient set at the above value and measure the chemotactic drift velocity. For our choice of parameters (see Table \ref{table}), $A= 5 \mu M$ and $\lambda = 200 \mu m$, we observe a drift velocity $ \simeq 1.5 \mu m/s$ which is much larger than the values shown in Fig. \ref{fig:v100}. Note that the drift velocity is known to be proportional to the gradient \cite{chatterjee2011chemotaxis, de2004chemotaxis} and when this gradient is set at the highest possible value, it is expected that the drift velocity will also be higher. 

\begin{figure}
\includegraphics[scale=0.7]{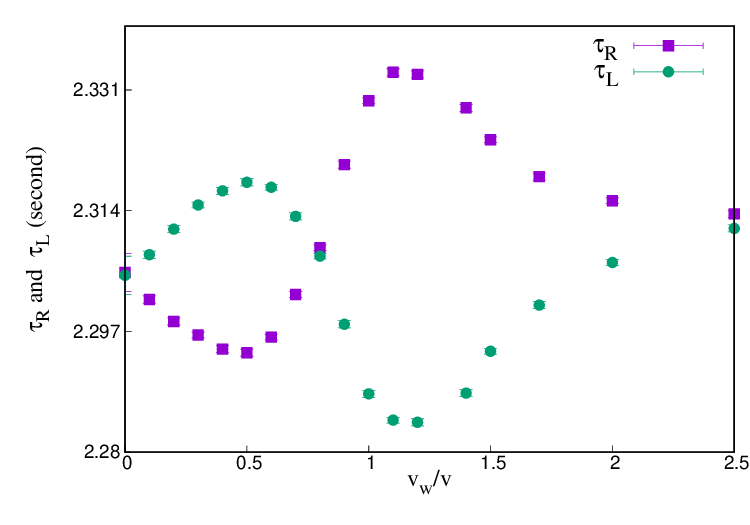}
\caption{Average rightward run duration $\tau_R$ (purple square points) and leftward run duration $\tau_L$ (green circular points) as a function of (scaled) wave speed. $\tau_R$ and $\tau_L$ show opposite trends as the wave speed is varied. The maximum error-bar in these data is $10^{-4}$ seconds. All simulation parameters are as in Fig. \ref{fig:v100}.}
\label{fig:trl100}
\end{figure}

To explain the nature of variation of $V$ in Fig. \ref{fig:v100}, we also separately look at the mean run durations in the same direction and opposite direction of wave propagation, denoted as $\tau_R$ (rightward) and $\tau_L$ (leftward), respectively. Since $V$ is measured from the net displacement in a run, the difference $(\tau_R - \tau_L)$ is closely related to the quantity $\Delta$ defined in Eq. \ref{eq:del} and can shed light on the behavior of $V$. As per our notation introduced at the beginning of this section, we can write  
\begin{equation}
\tau_R = \frac{\int dxd_R(x)}{\int dx N_R(x)}
\end{equation}
and 
\begin{equation}
 \tau_L = \frac{\int dx d_L(x)}{\int dx N_L(x)}.
\end{equation}
In Fig. \ref{fig:trl100} we show the variation of $\tau_R$ and $\tau_L$ with $v_w$. Our data show that for very small or very large $v_w$ both these run durations are equal. For $v_w =0$, when the environment is static, an uphill or a downhill run along the attractant concentration profile can result from both leftward and rightward movement of the cell. There is no net displacement of the cell towards left or right in the long term. On the other hand, when $v_w$ is very large, the cell can hardly sense the attractant gradient, rather it experiences an average attractant concentration $[L]_0$ at all times and can not therefore distinguish between a leftward and rightward run. So we have $\tau_R = \tau_L$ in both these limits which is consistent with our observation of $V=0$ in Fig. \ref{fig:v100}. For intermediate $v_w$ values, $\tau_R$ and $\tau_L$ show exactly opposite trends but it is clear from Fig. \ref{fig:trl100} that their difference $(\tau_R - \tau_L)$ indeed captures the $V$ variation in Fig. \ref{fig:v100} perfectly well. In the following subsection we focus on explaining the variation of $\tau_R$ and $\tau_L$ with $v_w$ in detail. 

As illustrated in Fig. \ref{fig:mov}, although the run speed of the cell is $v$, the effective velocity of a rightward run with respect to the moving wave is $(v-v_w)$ and that of a leftward run is $-(v+v_w)$. This means as $v_w$ increases from zero, rightward runs become slower and slower and finally when $v_w$ becomes equal to the run-speed, the cell simply rides the wave and experiences no change in attractant level during a rightward run. When $v_w$ exceeds $v$, the wave overtakes the right-movers, i.e. the cell effectively moves backward during a rightward run. For the left-movers, on the other hand, the effective velocity remains negative for all $v_w$ and its magnitude increases with $v_w$. The different attractant environment experienced by right-movers and left-movers plays an extremely important role in determining the variation of $\tau_R$ and $\tau_L$ with $v_w$, as illustrated in the next subsection.

\subsection{Runs originating from low and high attractant concentrations}
\label{sec:hilo}
\begin{figure}
\includegraphics[scale=2.5]{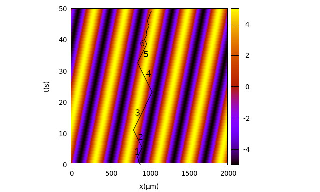}
\caption{ A typical run-and-tumble trajectory in a traveling wave attractant background. The high and low concentration regions are shown using the color scheme presented at the right, where the values correspond to $[L](x,t) - [L]_0$. Runs marked as $1$,$2$ and $4$ originate from low concentration regions. Out of these runs, $1$ contribute to $\tau_R^{low}$ and $2,4$ contribute to $\tau_L^{low}$. Similarly, runs $3$, $5$ are counted in $\tau_R^{high}$ and run $6$ in $\tau_L^{high}$. Here, we have used $\lambda = 400 \mu m$ and $v_w = 16 \mu m/s$.}
\label{fig:heat}
\end{figure}

We start by dividing all the runs into two groups: runs starting from regions with attractant concentration higher and lower than the average value $[L]_0$. Clearly, $\tau_R$ ($\tau_L$) is the average duration of all those rightward (leftward) runs, some of which have originated from high concentration regions, and the rest from low concentration regions. Runs which originate from a region with ligand concentration lower (higher) than $[L]_0$ can be both uphill or downhill, but majority of them will be uphill (downhill), i.e. the final concentration at the end of the run will be larger (smaller) than the initial concentration at its origin. This holds everywhere, except for the special case of rightward runs with $v_w=v$. We denote by $\tau_R^{low}$ and $\tau_R^{high}$ the mean duration of rightward runs starting from low and high concentration regions, respectively. Similarly, $\tau_L^{low}$ and $\tau_L^{high}$ can be defined. We investigate how these quantities depend on $v_w$. In Fig. \ref{fig:heat}  we show a typical run-and-tumble trajectory of the cell in the background of a traveling wave attractant environment. For the purpose of illustration, we mark a few representative runs originating from low and high attractant regions. In Appendix \ref{app:wake} we have presented data for runs starting from the rising and falling phases of the traveling wave.

\begin{figure}
\includegraphics[scale=1.5]{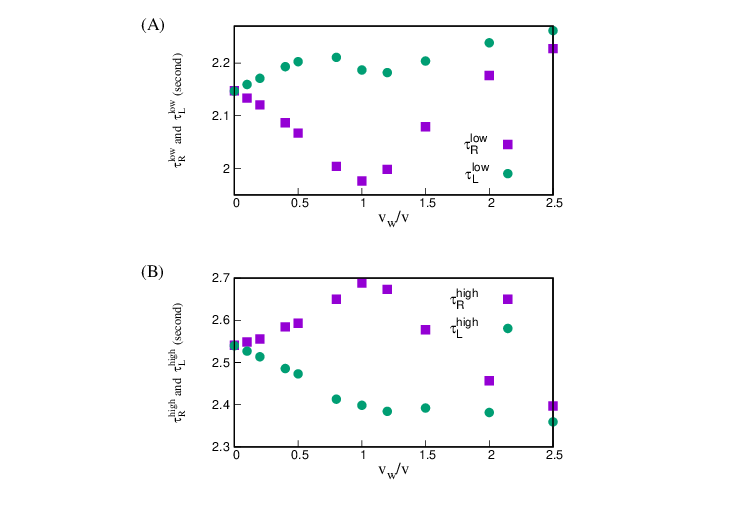}
\caption{(A) Average duration of rightward and leftward runs starting from regions with low attractant levels, as the wave speed is varied. When the wave speed equals the run speed, the right-movers feel no gradient. $\tau_R^{low}$ is minimum at this point. Left-movers experience steeper gradient throughout and $\tau_L^{low}$ shows an overall increasing trend with wave speed. (B) Similar quantities for high attractant levels show exactly opposite behavior from (A). The maximum error-bar in these data is $10^{-4}$ seconds. The simulation parameters are as in Fig. \ref{fig:v100}. }
\label{fig:minmax100} 
\end{figure}

Let us specifically consider the rightward runs which originate from a low concentration region. As $v_w$ increases from zero, the effective velocity $(v-v_w)$ of these runs decreases, i.e. the cell spends longer time in the low concentration region. Being exposed to low values of $[L]$ for a prolonged period of time raises its activity (see Eq. \ref{eq:nrg}) and makes it tumble quicker. Therefore $\tau_R^{low}$ should decrease with $v_w$, and reach a minimum at $v_w=v$, when attractant level remains unchanged during a run. For $v_w > v$ magnitude of effective velocity of a rightward run increases again, which makes it possible for some rightward runs to get out of the low concentration zone and enter the high concentration zone. This lowers the activity and increases the run duration. Our data (Fig. \ref{fig:minmax100}A, purple squares) indeed show that $\tau_R^{low}$ decreases with $v_w$, reaches a minimum at $v_w=v$ and then increases again. The behavior of $\tau_R^{high}$ is just the opposite (Fig. \ref{fig:minmax100}B, purple squares), and can be explained similarly: for $0 < v_w < v$ the relative velocity of the right-movers keeps decreasing with $v_w$, which makes the cell spend longer time in the high concentration region and hence $\tau_R^{high}$ increases with $v_w$, reaching a maximum at $v_w=v$ and then decreases again. 

Final $\tau_R$ is calculated after averaging over all runs, starting from low and high concentration zones. In Fig. \ref{fig:mimaf} we plot the total number of these runs for different values of $v_w$. Since more tumbles occur in the low concentration region, more runs originate from here (verified in Fig. \ref{fig:mimaf}). The behavior of $\tau_R$ is therefore initially controlled by $\tau_R^{low}$ and $\tau_R$ decreases with $v_w$ (Fig. \ref{fig:trl100}, purple squares). But when $\tau_R^{low}$ falls much below $\tau_R^{high}$, its contribution to $\tau_R$ becomes less significant and $\tau_R^{high}$ takes over. Therefore, $\tau_R$ increases again, reaches a peak and finally for very large $v_w$ merges with $\tau_L$.

For leftward runs, effective velocity keeps increasing with $v_w$. This means when a cell is moving leftward, the effective concentration gradient experienced by the cell keeps increasing with $v_w$. A steeper gradient elongates the uphill runs and shortens the downhill runs. Since $\tau_L^{low}$ ($\tau_L^{high}$) is obtained by averaging over majority uphill (downhill) runs, $\tau_L^{low}$ ($\tau_L^{high}$) shows increasing (decreasing) trends with $v_w$. Our data in Fig. \ref{fig:minmax100} (green circles) are consistent with this, except for $v_w \sim v$, where these trends are slightly disrupted. This effect can be explained in the following way. When wave propagation speed is close to the run speed of the cell, $\tau_R^{low}$  is particularly small, since these runs are associated with high activity. When these rightward runs tumble, half of them turn leftward and contribute towards $\tau_L^{low}$. Such leftward runs therefore start with high activity and are more likely to tumble quickly. This brings down the average $\tau_L^{low}$ near $v_w=v$ point. In a similar way, mild increasing tendency of $\tau_L^{high}$ for $v_w \sim v$ can also be explained.

$\tau_L$ is the weighted average of $\tau_L^{high}$ and $\tau_L^{low}$. Due to higher tumbling rate in the low concentration region, more runs tend to originate from there. So $\tau_L$ behavior is mainly guided by $\tau_L^{low}$: it increases for small and large $v_w$ and decreases when $v_w$ is close to $v$. Although $\tau_L^{high}$ does contribute to the variation of $\tau_L$, it never takes over from $\tau_L^{low}$ because the difference between high and low concentration run durations for left-movers is much less compared to that for right-movers.


\section{Chemotactic response for large wavelength} \label{sec:large}

\begin{figure}
\includegraphics[scale=0.7]{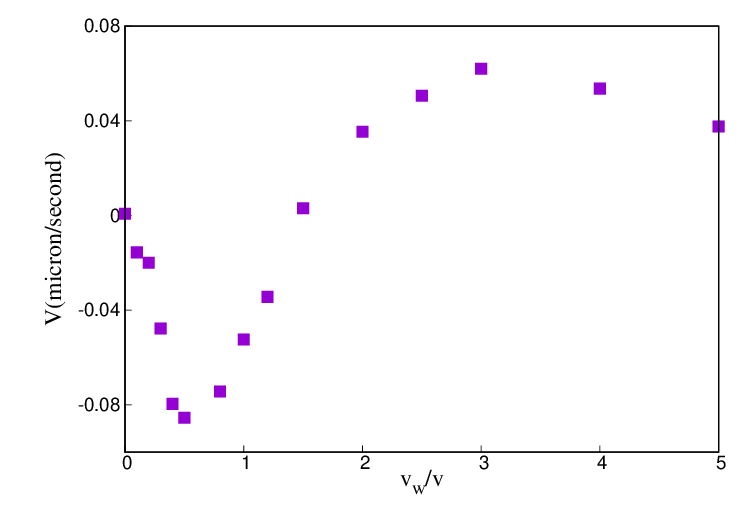}
\caption{Variation of drift velocity $V$ with (scaled) wave speed $v_w$ for large wavelength. The qualitative behavior is same as that for moderate $\lambda$ values. The maximum error-bar in these data is $10^{-3} \mu m/s$.Here, we have used $\lambda = 1000 \mu m$ here and all other simulation parameters are as in Fig. \ref{fig:v100}.} \label{fig:v500}
\end{figure}

\begin{figure}
\includegraphics[scale=0.7]{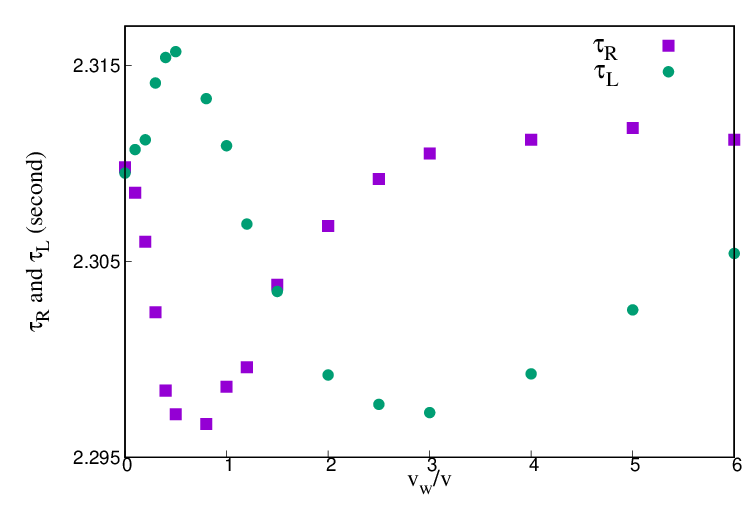}
\caption{Average duration of rightward and leftward runs as a function of scaled wave speed for large wavelength. The difference between these two run durations correctly capture the behavior of $V$ depicted in Fig. \ref{fig:v500}. The maximum error-bar in these data is $10^{-4} s$. All simulation parameters are as in Fig. \ref{fig:v500}. }
\label{fig:trl500}
\end{figure}

The qualitative variation of $V$ with $v_w$ remains similar even for large $\lambda$ (Fig. \ref{fig:v500}) although variation in $V$ occurs over a much smaller range in this case. This is not surprising, since for larger $\lambda$, the attractant concentration gradient is smaller. Our data for $\tau_R$ and $\tau_L$ in Fig. \ref{fig:trl500} show that qualitative dependence of mean run durations on $v_w$ remains similar too, although the positions of maxima and minima get shifted. However, when we separately measure runs originating from high and low concentrations of attractant, we find different behaviors. Although $\tau_R^{low}$ ($\tau_R^{high}$) shows qualitatively similar behavior with $v_w$, as in Fig. \ref{fig:minmax100}, the minimum (maximum) is reached at a $v_w > v$. More importantly, the qualitative trends of $\tau_L^{low}$ and $\tau_L^{high}$ are quite different. $\tau_L^{low}$ shows a minimum with $v_w$ and $\tau_L^{high}$ shows a maximum. 
\begin{figure}
\includegraphics[scale=1.5]{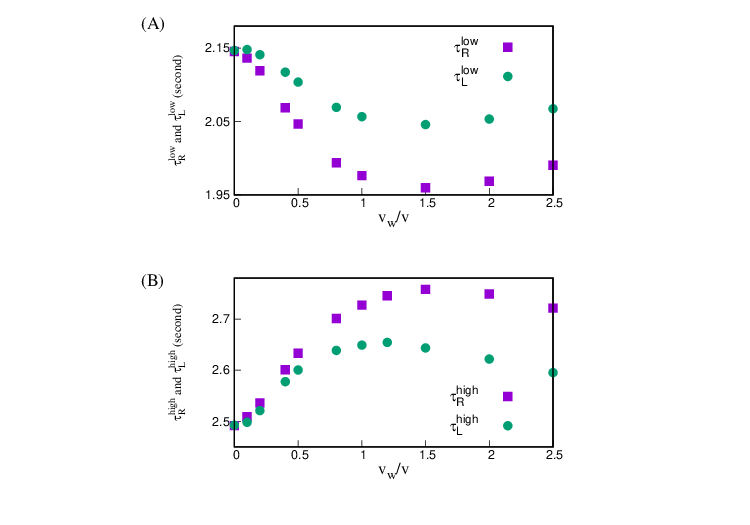}
\caption{(A) Mean duration of rightward and leftward runs originating from regions with low attractant levels, when the wavelength is large. The qualitative behavior is markedly different in this case from moderate wave length. While $\tau_R^{low}$ shows a minimum with $v_w$, its position is shifted from $v_w=v$ point. Moreover, $\tau_L^{low}$ also shows similar nature of variation as $\tau_R^{low}$, unlike the case for moderate $\lambda$. (B) Similar quantities for runs originating from high attractant level. The maximum error-bar in these data is $10^{-4}$ seconds. All simulation parameters are as in Fig. \ref{fig:v500}. }
\label{fig:minmax500}
\end{figure}

To explain this observation, we note that for very large $\lambda$, the spatial variation of ligand concentration is so slow, that even when $v_w$ is not equal to $v$, the change in ligand level experienced by the cell during a rightward run is negligible. Therefore, the dramatic effect observed for $v_w=v$ when $\lambda$ was small (Fig. \ref{fig:minmax100}) is much subdued here. The low concentration zone, which spans over half a wavelength, is so large in this case, that many of our arguments for smaller $\lambda$ do not remain valid here. For example, our explanation behind $\tau_L^{low}$ variation presented in Fig. \ref{fig:minmax100}A hinged on the fact that the majority of the leftward runs originating from low concentration regions are uphill, and these runs control the average. But when the low concentration region is spanned over a very large spatial range, then the number of downhill runs originating from this region is comparable and is expected to affect the behavior of $\tau_L^{low}$.

\begin{figure}
\includegraphics[scale=0.7]{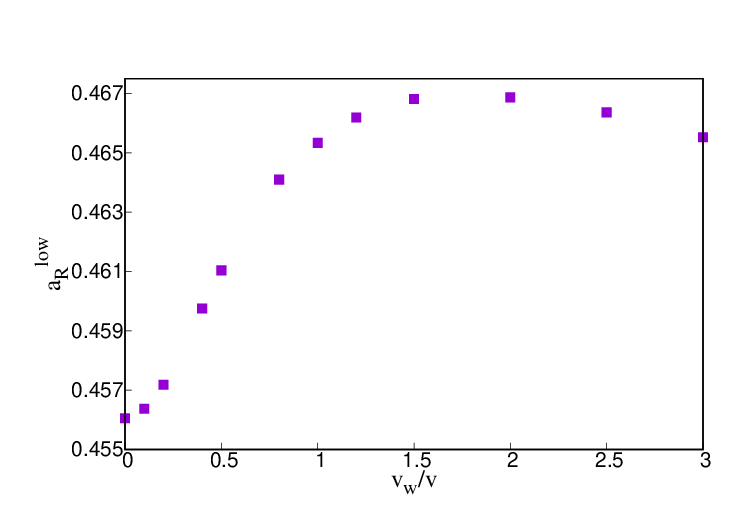}
\caption{Initial activity of a rightward run, originating from low attractant regions, as a function of wave speed. In the case of large wavelength, {\sl i.e.} weak attractant gradient, initial activity effectively determines the run duration. Behavior of $a_R^{low}$ is consistent with that of $\tau_R^{low}$ in Fig. \ref{fig:minmax500}A. The maximum error-bar for these data is $10^{-5}$. All simulation parameters are as in Fig. \ref{fig:v500}.} 
\label{fig:a0tr}
\end{figure}

For large $\lambda$, the change in ligand level and hence change in activity is so small during a typical run, that the duration of a run is effectively determined by the initial activity of the cell at the beginning of that run. In Fig. \ref{fig:a0tr} we plot the average initial activity of rightward runs starting from low concentration regions. We denote this quantity by $a_R^{low}$. We find that $a_R^{low}$ indeed shows a peak at the same $v_w$ where $\tau_R^{low}$ is minimum. Since direction of a run is randomized after each tumble, same behavior is expected even for $a_L^{low}$. This is why $\tau_R^{low}$ and $\tau_L^{low}$ show the same qualitative trend in this case. In a similar way, data in Fig. \ref{fig:minmax500}B can also be explained. However, we do not have complete understanding of why $a_R^{low}$ in Fig. \ref{fig:a0tr} shows a peak with $v_w$ and why that peak occurs at $v_w > v$. More research is needed to understand these details. In Appendix \ref{app:a0zone} Fig. \ref{fig:acwhol} we also present data for initial activity for few more concentration ranges of attractant and in all cases find behavior consistent with Fig. \ref{fig:minmax500}. Note that initial activity correctly captures the variation of mean run durations only when $\lambda$ is very large. For moderate $\lambda$ values the variation in activity during a typical run is significantly higher, and the initial activity does not control anymore when the cell is going to tumble next. In Appendix \ref{app:a0zone} Fig. \ref{fig:acwhol100} we explicitly show this.  

\section{Conclusions}
\label{sec:con}
In summary, we have considered the chemotaxis of a single {\sl E.coli} cell in a traveling wave attractant environment. We show that the steady state chemotactic drift velocity of the cell, which is measured by averaging over displacement of the cell at all times, shows non-trivial dependence on the wave speed with multiple peaks and zero-crossings. We argue that this behavior arises due to very different attractant gradients experienced by the cell when it is moving in the same or opposite direction of wave propagation. We also find significant difference in cell response for moderate and large wavelength.

Our study highlights the importance of performing dc measurements even when the steady state is time-periodic in nature. In another related study \cite{li2017barrier} chemotaxis was considered for a bacterial population in presence of a traveling wave attractant profile. But in this case drift velocity was measured from the forward and backward movement of the cell population within one time period of the wave. This is distinctly different from our definition where averaging over all times is performed, irrespective of the phase of the wave. As defined in Eqs. \ref{eq:tau}, \ref{eq:del} and \ref{eq.vdef}, we measure net displacement of the cell during a run, and average over a large number of runs. In Appendix \ref{app:u} we use another definition of drift velocity, where we measure the net displacement in a time-interval much longer than the typical run duration and average over many such intervals. In both these definitions, we need to consider really long steady state trajectories of the cell, which extend across many time-periods. Indeed comparison with \cite{li2017barrier} shows that this is an important difference which completely changes the qualitative nature of variation. While we find negative $V$ for small $v_w$, in \cite{li2017barrier} a positive drift velocity was reported. Apart from this major difference, there are also other differences related to the attractant environment considered in \cite{li2017barrier} where a traveling wave is generated by solving three dimensional diffusion equation with periodically active source terms. This yields a traveling wave form which is much more complex than the simple sinusoidal wave we consider here. Specifically, at different time instants the shapes of the attractant profile prevailing in the channel are very different. In addition, the amplitude of the traveling wave is significantly higher, and the propagation speed typically lower in \cite{li2017barrier} compared to our present study. These factors also play a role in determining the drift velocity.

The presence of a traveling wave stimulus profile causes chemotactic response in the system which also has similar traveling wave form, in the long time limit. Indeed we find the cell density $P(x,t)$, for example, is a function of $(x-v_w t)$, and propagates through the system just like $[L](x,t)$ in Eq. \ref{eq:cxt}. However, there is a phase lag between $P(x,t)$ and $[L](x,t)$ which increases with wave speed (data in Fig. \ref{fig:lag} in Appendix \ref{app:lag}). This is consistent with the expectation that $P(x,t)$ tries to follow the moving attractant wave and as the wave moves faster, $P(x,t)$ starts lagging behind. Our data in Fig. \ref{fig:lag} show that when the wave speed is close to the run speed of the cell, the phase lag is $\sim \pi$, which means the maximum of $P(x,t)$ occurs near the minimum of $[L](x,t)$ (Fig. \ref{fig:pxcxlg}B and Fig. \ref{fig:pxcxlgL}B). For large wave speed, we record a phase difference $\sim 2 \pi$, {\sl i.e.} the peak in $P(x,t)$ now moves to the next attractant peak. A similar effect was reported in \cite{li2017barrier} where it was observed experimentally that for a fast attractant wave, the cell cluster moves to the next peak of the traveling wave attractant profile. Even in absence of any spatial periodicity, a phase lag was observed when a time-periodic attractant environment was created inside a microfluidic device \cite{zhu2012frequency}. It was shown that for a large time-period, the temporal variation of bacterial population happens in sync with the attractant wave, but for a  smaller time-period, $T \sim 100s$ there is a phase lag of $\pi$. Our results in Fig. \ref{fig:lag} suggest that for even smaller $T$, the phase lag should exceed $\pi$ and it would be interesting to check this prediction against experiments.

Throughout this work, we have performed simulations for an attractant concentration that varies sinusoidally in space and time. However, our conclusions should remain valid for any other traveling wave profile. Our explanation behind the variation of chemotactic drift velocity with the wave speed hinges on the fact that in the same and opposite  direction of wave propagation, different attractant gradients are faced by the cell. This is true for any kind of traveling wave pattern, irrespective of its shape. This generalization is particularly significant for possible experimental realization of our results. While it might be challenging to set up a perfect sine wave moving with a constant speed, other traveling wave forms have been possible to generate in microfluidic channels \cite{zhu2012frequency, lazova2011response, li2017barrier}. Often in experiments microfluidic channels are considered with width comparable to the typical run length of the cell \cite{li2017barrier} and in this case, the effective motion of the cell remains one dimensional, as was considered here in our model. However, we have verified (data in Appendix \ref{app:2d}, Fig. \ref{fig:v1002d}) that even when run-and-tumble motion takes place in two dimensions, a traveling wave attractant profile generates a similar response in the drift motion. This observation may be relevant for possible experimental verification of our results.

Beyond bacterial chemotaxis, and in a more general context, our study shows a new mechanism for generating negative drift velocity. Chemical gradient is the reason behind drift motion in our system, and by ensuring that the cell faces very different gradients while moving in different directions, we find negative drift velocity. This mechanism is distinctly different from other active systems \cite{Ghosh2014, Rizkallah2023, straube23} where negative mobility has been observed. For example, in \cite{Ghosh2014} transport of  active Janus particles through a narrow corrugated channel was numerically studied and transport in the opposite direction of applied force was reported. This effect was shown to be caused by frequent tumbling (or effective trapping) of the Janus particles at the wall protrusions and was also found to be sensitive to the shape of the particles. The effect was most pronounced when the particles were rod-like with active motion directed along the body axis, in which case giant absolute negative mobility was found. In \cite{Rizkallah2023} absolute negative mobility was observed for an active tracer moving in an environment of passive crowders with excluded volume interactions. It was analytically shown that for small external force, and large persistence time of the active particle, the trapping effect caused by the passive crowders can give rise to absolute negative mobility. Compared to these systems, the mechanism we propose here is qualitatively different and it will be of general interest to find other active systems where our mechanism can be used to explain negative mobility.

\section{Acknowledgements}
SC acknowledges support from Anusandhan National Research Foundation (ANRF), India (Grant No: CRG/2023/000159).

\appendix 
\section{Additional model details} 
\label{app:model}

Number of CheR and CheB molecules are of the order of hundred, compared to few thousand receptor dimers present in the cell (see Table \ref{table}). To explain how so few enzyme molecules manage to regulate methylation levels of all the receptors and still maintain robust adaptation \cite{berg1975transient, goy1977sensory}, few mechanisms like assistance neighborhood and brachiation have been suggested and experimentally verified  \cite{hansen2008chemotaxis, levin2002binding, endres2006precise, li2005adaptational, kim2002dynamic}. In assistance neighborhood, an enzyme molecule, which is bound to one particular dimer can also modify the methylation levels of neighboring dimers, while in brachiation a bound enzyme performs a random walk on the receptor array and modify the methylation levels of all dimers along its trajectory. In our simple model, we have included a flavor of both these mechanisms in the binding-unbinding kinetics of the enzymes.

A free CheR molecule in the cell cytoplasm can bind to an unoccupied dimer with a small rate $w_r$ \cite{feng1999enhanced, pontius2013adaptation, wu1996receptor, pontius2013adaptation, schulmeister2008protein}. Once it is bound, the CheR enzyme can raise the methylation level of the dimer by one unit with a rate $k_r$ if the dimer belongs to an inactive cluster and methylation level of the dimer is less than $8$. A bound CheR molecule can also unbind from the dimer with a rate $w_u$ and rebind with rate $w_{rb}$ to another randomly chosen dimer within the same cluster, provided it is unoccupied. If the rebinding attempt is unsuccessful, the CheR molecule then returns to the cytoplasm. This rebinding process is much faster compared to the binding of a CheR in the cytoplasm. Such multiple rebinding events within same cluster ensure that multiple methylation reactions can take place before the enzyme returns to the cytoplasm. Using this mechanism we include the flavour of assistance neighbourhood \cite{endres2006precise,hansen2008chemotaxis, li2005adaptational} and enzyme brachiation \cite{levin2002binding} mechanisms in our simple model.

A CheB molecule in the cytoplasm can get phosphorylated by an active receptor with a rate $w_p$  and dephosphorylation of an unbound CheB-P can happen with a rate $w_{dp}$. The binding-unbinding rules of a CheB-P molecule are similar to what we have used for CheR. From the cell cytoplasm, a CheB-P molecule can bind to the tether site of a dimer with rate $w_b$, provided no other enzyme is already bound to the dimer. After binding the enzyme can decrease the methylation level of the dimer with a rate $k_b$ by one unit provided the dimer belongs to an active cluster and its methylation level has a non-zero value. The unbinding and rebinding processes of CheB-P happen in similar way as explained in the previous paragraph.

\begin{figure}
\includegraphics[scale=1]{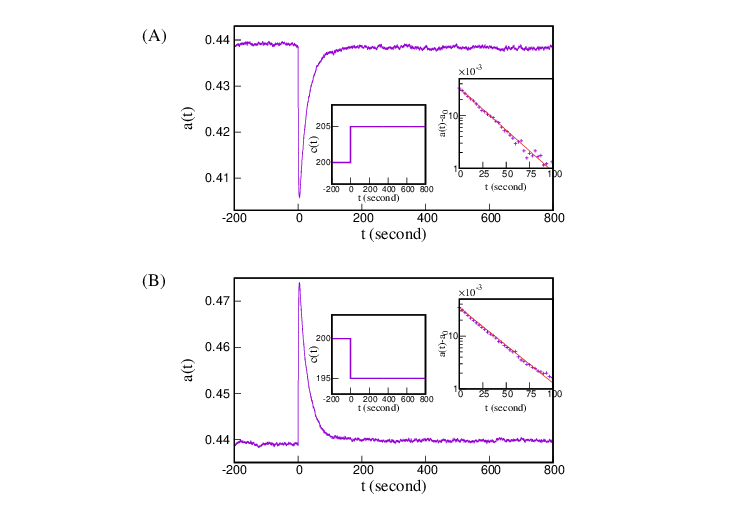}
\caption{Time variation of activity in response to a step up and step down stimulus. In both cases, in the long time limit, the activity relaxes exponentially with a time-scale $\tau_m \sim 27.9s$. The step size used here is $5 \mu M$. All other simulation parameters are as in Fig. \ref{fig:v100}.}
\label{fig:stpud}
\end{figure}

We have used the above model of (un)binding kinetics of enzyme molecules and (de)methylation of receptors in earlier works to study in detail the methylation dynamics for both tethered cell and swimming cell \cite{mandal2022effect, chatterjee2022short}. In Fig. \ref{fig:stpud} we plot temporal variation of activity in response to step up and step down stimulus. We use a step size of $5$ $\mu M$ which is same as the value of $A$ in Eq. \ref{eq:cxt}. We measure adaptation time-scale $\tau_m \sim 27.9s$.        

\subsection{Simulation Algorithm}
We present the values of all model parameters in Table \ref{table}. In our simulation we use the following algorithm to implement the model for signaling pathway and run-tumble motion of the cell:
\begin{itemize}
\item {\bf Activity switching}: For each receptor cluster calculate $F([L],m)$ from Eq. \ref{eq:nrg}. If the cluster is in the inactive state, then make it active with probability $\dfrac{w_a dt}{1+exp(F)}$. Or if the cluster is in the active state, with probability $\dfrac{w_a dt}{1+exp(-F)}$ it switches to inactive state. 

\item {\bf Phosphorylation-dephosphorylation of CheB}: If a CheB molecule is in the unphosphorylated state, then it gets phosphorylated with probability $aw_p dt$, where $a$ is the fraction of active receptor clusters at that time. If the CheB molecule is already phosphorylated, and if it is not bound to any receptor, then it gets dephosphorylated with probability $w_{dp}dt$. We do not allow dephosphorylation of bound CheB-P.

\item {\bf Binding-unbinding dynamics of CheR and methylation}: If a CheR molecule is in the unbound state in cell cytoplasm, then bind it to a randomly selected receptor dimer with probability $w_r dt$ provided the chosen dimer is not already bound to another enzyme molecule. On the other hand, if the CheR molecule is already bound to a dimer, and if the dimer is in the inactive state with methylation level $m <8$, then with probability $k_r dt$ its methylation level is raised by unity. At the end of the modification attempt, the bound CheR molecule unbinds from the dimer with probability $w_u dt$ and rebinds to another randomly selected dimer from the same cluster, provided no other enzyme molecule is bound to it. The rebinding process happens with rate $w_{rb}$ which is very high (see Table \ref{table}). For our choice of $dt=0.01s$, we assign probability $1$ to a rebinding event. If the rebinding attempt is unsuccessful (because the selected dimer was occupied), then the CheR molecule returns to cell cytoplasm.  

\item {\bf Binding-unbinding dynamics of CheB-P and demethylation}: Follows very similar algorithm as described in the previous paragraph. Only $w_r$ is replaced by $w_b$, $k_r$ is replaced by $k_b$ and the demethylation condition changes to active dimer with $m >0$. All other steps remains same. 

\item {\bf Switching between run and tumble modes}: Calculate $G=\Delta_1 - \dfrac{\Delta_2}{1+Y_0/Y_P} $ where $Y_P = \dfrac{a}{a+K_Z/K_Y} $ denotes the fraction of phosphorylated CheY molecules and $a$ is the fraction of active receptor clusters. If the cell is in the run mode, then with probability $\omega dt \exp(-G)$ it switches to tumble mode. If the switching attempt is not successful, then the cell continues to run. If, on the other hand, the cell is in tumble mode, then with probability $\omega dt \exp(G)$ it switches to run mode.

In one dimension, while the cell is running, its position is updated as $x(t+dt)=x(t)+vdt$ where $v$ is the run speed. We also apply periodic boundary condition here, such that $x(t)$ always remains bounded between $0$ and $L$. After each successful transition from tumble to run mode, a random sign for $v$ is chosen which does not change until the next tumble. 

In two dimensions, during a run, the cell position is updated as $x(t+dt)=x(t)+vdt \cos \theta (t)$ and $y(t+dt)=x(t)+vdt \sin \theta (t)$, with periodic boundary condition in both $x$ and $y$ directions. The angle $\theta (t)$ is obtained from numerically integrating $\dfrac{d \theta }{dt} = \eta (t)$. After each successful transition from tumble to run mode, a random value of $\theta$ is chosen between $0$ and $2 \pi$. 

\item Go back to activity switching step and repeat the whole process.

\end{itemize}

\begin{center}
\begin{table}
\caption{List of model parameters}
\begin{tabular}{|l|l|l|l|}
\hline
\textbf{Symbol} & \hspace{30mm} \textbf{Description} & \textbf{Value} & \textbf{References} \\
\hline
$n$ & number of trimers of dimers in a cluster & $10$ & present study \\
\hline
$N_{dim}$ & Total number of receptor dimers & $7200$ & \cite{pontius2013adaptation,li2004cellular} \\
\hline
$N_r$  & Total number of CheR  protein molecules &  $140$ & \cite{pontius2013adaptation,li2004cellular}\\
\hline
$N_b$  & Total number of CheB  protein molecules &  $240$ & \cite{pontius2013adaptation,li2004cellular} \\
\hline
$\epsilon_0$ & Basal energy of receptor dimer & $1$ $k_B T$ & \cite{pontius2013adaptation, frankel2014adaptability, dufour2014limits, long2017feedback} \\
\hline
 $\epsilon_1$ & Receptor energy change per methyl group addition & $1$ $k_B T$  & \cite{pontius2013adaptation,frankel2014adaptability,dufour2014limits,long2017feedback} \\
\hline
$K_{min}$ & Ligand dissociation constant for inactive receptor &  $18$ $\mu M$ & \cite{jiang2010quantitative}, \cite{flores2012signaling} \\
\hline
$K_{max}$ & Ligand dissociation constant for active receptor &  $3000$ $\mu M$ & \cite{flores2012signaling,jiang2010quantitative} \\
\hline
$w_{a}$  & Flipping rate of activity &  $0.75$ $s^{-1}$ & Present study\\
\hline
$\omega$  & Switching frequency of motor &  $1.3$ $s^{-1}$ & \cite{sneddon2012stochastic, sneddon2011efficient} \\
\hline
 $\Delta_{1}$  & Nondimensional constant regulating motor switching  &  $10$ & \cite{sneddon2012stochastic, sneddon2011efficient} \\
\hline
$\Delta_{2}$  & Nondimensional constant regulating motor switching &  $20$ & \cite{sneddon2012stochastic, sneddon2011efficient}  \\
\hline
$Y_{0}$ & Adopted value of the fraction of CheY-P protein &  $0.34$ & \cite{sneddon2012stochastic, sneddon2011efficient} \\
\hline 
$K_{Y}$ & Phosphorylation rate of CheY molecule &  $1.7$ $s^{-1}$ & \cite{flores2012signaling,tu2008modeling} \\
\hline
$K_{Z}$ & Dephosphorylation rate of CheY molecule &  $2$ $s^{-1}$ & \cite{flores2012signaling,tu2008modeling}\\
\hline
$w_{r}$ & Binding rate of bulk CheR to tether site of an unoccupied dimer &  $0.068$ $s^{-1}$ & \cite{pontius2013adaptation,schulmeister2008protein} \\
\hline
$w_{b}$ & Binding rate of bulk CheB-P to tether site of an unoccupied dimer &  $0.061$ $s^{-1}$ & \cite{pontius2013adaptation,schulmeister2008protein} \\
\hline
$w_{u}$ & Unbinding rate of bound CheR and CheB-P &  $5$ $s^{-1}$ & \cite{pontius2013adaptation,schulmeister2008protein} \\	
\hline
$w_{rb}$ & Rebinding rate of CheR and CheB-P to another dimer & $1000$ $s^{-1}$ & \cite{pontius2013adaptation} \\
\hline	
$k_{r}$ & Methylation rate of bound CheR & $2.7$ $s^{-1}$ & \cite{pontius2013adaptation,schulmeister2008protein}\\ 	
\hline
$k_{b}$ & Demethylation rate of bound CheB-P &  $3$ $s^{-1}$ & \cite{pontius2013adaptation,schulmeister2008protein}\\
\hline	
$w_{p}$ & CheB phosphorylation rate &  $3$ $s^{-1}$ & \cite{pontius2013adaptation,stewart2000rapid} \\
\hline
$w_{dp}$ & CheB-P dephosphorylation rate &  $0.37$ $s^{-1}$ & \cite{pontius2013adaptation}\\	
\hline	
$L$ &  Channel length for $1d$ run-tumble motion &  $2000$ $\mu m$ & Present study \\
\hline
$L_x \times L_y$ & System size for $2d$ run-tumble motion & $2000 \mu m \times 800$ ${\mu m}$ & present study  \\
\hline
$v$ & Speed of the cell during a run &  $20$ $\mu m/s$ & \cite{2008cbergoli} \\
\hline
$dt$ & Time step &  $0.01$ $s$ & Present study\\
\hline
$D_\theta$ & Rotational diffusivity & $0.062$ $rad^2/s$ & \cite{berg1972chemotaxis, dufour2014limits} \\ 
\hline
$[L]_{0}$ & Background attractant concentration &  $200$ $\mu M$ & Present study \\
\hline
$A$ & Amplitude of the sine wave attractant profile &  $5$ $\mu M$ ($1d$),  $10$ $\mu M$ ($2d$) & Present study \\
\hline
$\lambda$ & Wave length of the sinusoidal attractant profile & $200$ $\mu m$, $1000$ $\mu m$ & Present study \\
\hline
\end{tabular}
\label{table}
\end{table}
\end{center}


\section{Data for two dimensions} \label{app:2d}

 For run-tumble motion taking place in $2d$ we consider an area of size $L_x \times L_y$ with periodic boundary conditions in both directions. The attractant gradient is applied only along $x$-direction, as in Eq. \ref{eq:cxt}. During a run, the cell makes an angle $\theta$ with the $x$-axis and at each tumble it chooses any value of $\theta$ from the range $[0,2 \pi]$ with uniform probability. Moreover, during a run $\theta$ does not remain constant, but it undergoes (slow) diffusion with diffusivity $D_\theta$ whose value is listed in Table \ref{table}. To implement this in our simulations, we numerically integrate the stochastic differential equation $\dfrac{d \theta}{dt} = \eta (t) $ using Euler–Maruyama method \cite{VANKAMPEN2007396}. Here $\eta (t)$ is Gaussian white noise which satisfies $\langle \eta (t) \eta (t') \rangle = 2 D_\theta \delta (t-t')$. At each time step the position $(x,y)$ of the cell is then updated as $x(t+dt) = x(t)+vdt \cos \theta(t) $ and $y(t+dt) = y(t)+vdt \sin \theta(t) $. Thus during a run the trajectory does not remain a straight line but bends gradually, as seen in experiments. 

\begin{figure}
\includegraphics[scale=0.7]{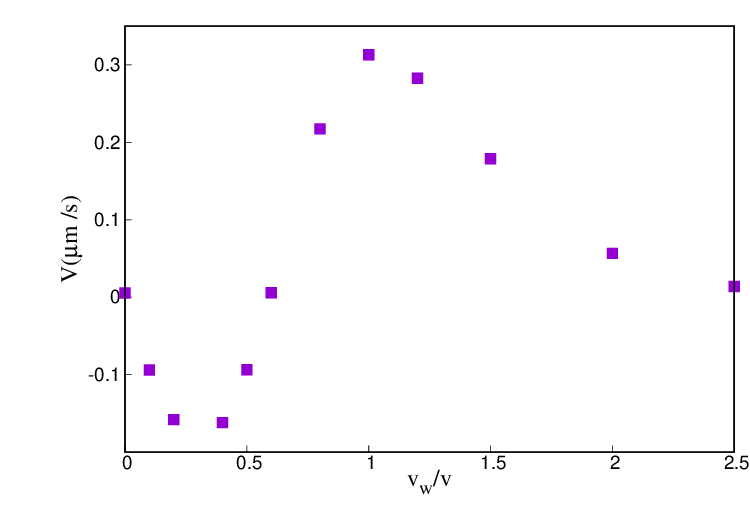}
\caption{Variation of chemotactic drift velocity $V$ with propagation speed $v_w$ of the attractant wave in two dimensions. The maximum error-bar in our measurement is $0.001 \mu m /s$. These data are for $\lambda = 200 \mu m$, $A=10 \mu M$. All other simulation parameters are listed in Table \ref{table}. }
\label{fig:v1002d}
\end{figure}

To calculate the drift velocity $V$ in $2d$ we observe the system for a long time $\mathcal T$ after reaching the steady state. Since drift motion is expected only along $x$ direction, we still define $V$ in terms of rightward and leftward runs. The displacement along $x$ direction in a single time-step is $v\cos \theta dt$ and we sum over all these time-steps during a run. If the sum is positive (negative), we classify it as a rightward (leftward) run. Then we calculate $V$ following the definition given in Eq. \ref{eq.vdef}. In Fig. \ref{fig:v1002d} we present our data for $V$ in two dimensions for different values of wave speed and find similar qualitative nature of variation as found in $1d$.   

\section{Mean displacement per unit time} 
\label{app:u}

\begin{figure}
\includegraphics[scale=0.7]{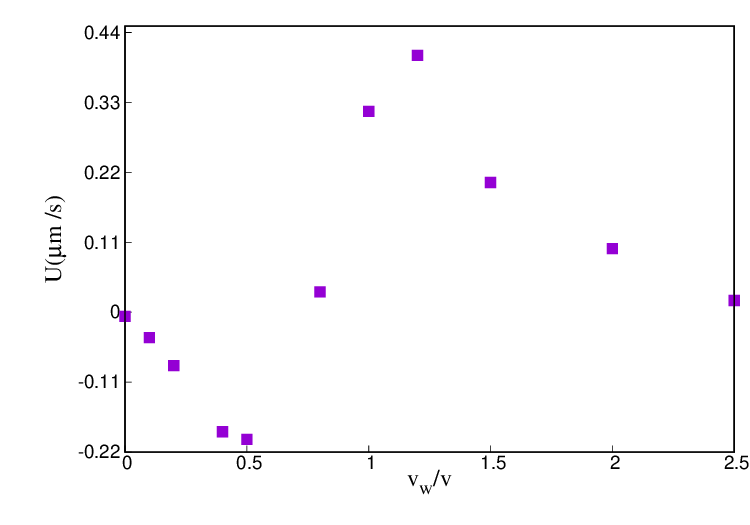}
\caption{Variation of drift velocity, measured as mean displacement per unit time, as a function of scaled wave speed. The nature of variation is same as in Fig. \ref{fig:v100}. All parameters are as in Fig. \ref{fig:v100}.}  \label{fig:u}
\end{figure}

In Fig. \ref{fig:v100} we have defined the chemotactic drift velocity as mean displacement in a run, divided by mean run duration. Another simpler definition of drift velocity is mean displacement per unit time. In steady state we observe the system for a time interval $\cal T$ (much longer than typical run duration) and measure the displacement in that interval. We average over many such intervals to calculate the mean displacement. After dividing this mean displacement by $\cal T$ we get another measure of drift velocity, which we denote as $U$. In Fig. \ref{fig:u} we plot $U$ as a function of wave speed and find the same behavior as $V$. These data are for $\lambda = 200 \mu m$. 

\section{Fraction of runs starting from low and high concentration zones }
\label{app:frac}

\begin{figure}
\includegraphics[scale=0.7]{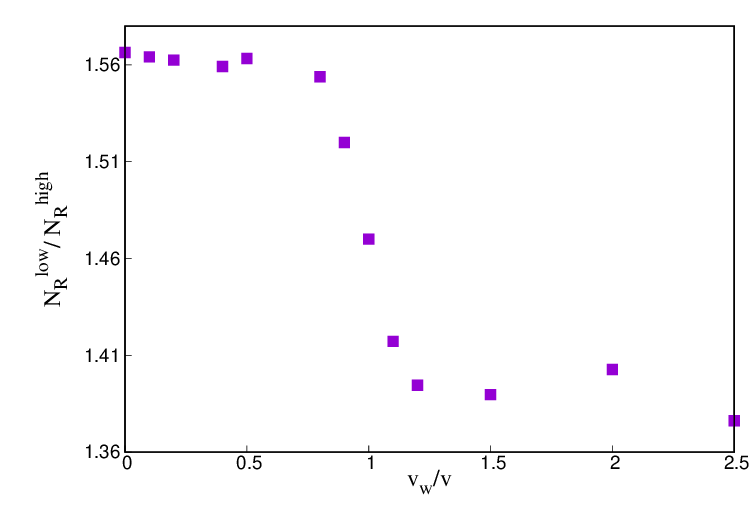}
\caption{Plot of the ratio $N_R^{low}/N_R^{high}$ for different wave speed. The simulation parameters are as in Fig. \ref{fig:v100}. }
\label{fig:mimaf} 
\end{figure}

Let $N_R^{low}$ denote the number of rightward runs starting from regions with low attractant concentration and $N_R^{high}$ denote the same for high attractant concentration. In Fig. \ref{fig:mimaf} we plot a ratio of these two quantities as a function of the scaled wave speed. Since more runs originate from low concentration zones, the ratio always remains larger than $1$. Analogous quantity for leftward runs also show identical behavior since run directions are randomized after each tumble. 

\section{Phase lag between $P(x,t)$ and $[L](x,t)$} \label{app:lag}

\begin{figure}
\includegraphics[scale=0.8]{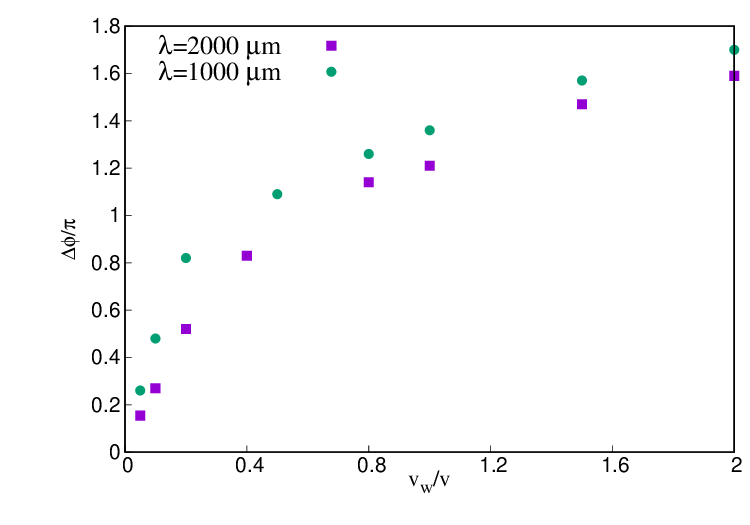}
\caption{Phase lag $\Delta \phi$ (measured in units of $\pi$) between $[L](x,t)$ and $P(x,t)$ as a function of the scaled wave speed. As wave moves faster, $P(x,t)$ starts lagging behind $[L](x,t$ more and more. $\lambda$ values are given in the legends. The maximum error-bar in these data is $0.01$. All other simulation parameters are as in Fig. \ref{fig:v500}.}
\label{fig:lag}
\end{figure}

For a static attractant profile, it is most likely to find the cell near the peaks of the attractant concentration. For a propagating traveling wave, the cell density still tries to catch up with the concentration peaks and follows the wave. However, as the propagation speed increases, the cell density starts lagging behind. In this appendix we measure this lag. To be more specific, for a traveling wave attractant profile, the position distribution $P(x,t)$ of the cell also has a traveling wave form in the long time limit, with the same spatial and temporal periodicity as $[L](x,t)$. The phase difference $\Delta \phi$ between $P(x,t)$ and $[L](x,t)$ has a finite value in steady state. In Fig. \ref{fig:lag} we plot this phase difference with $v_w/v$  for two different $\lambda$ values. We choose particularly large $\lambda$ values here since for smaller $\lambda$, there are many peaks of $[L](x,t)$ and $P(x,t)$ when it becomes difficult to track them. 

\begin{figure}
\includegraphics[scale=1]{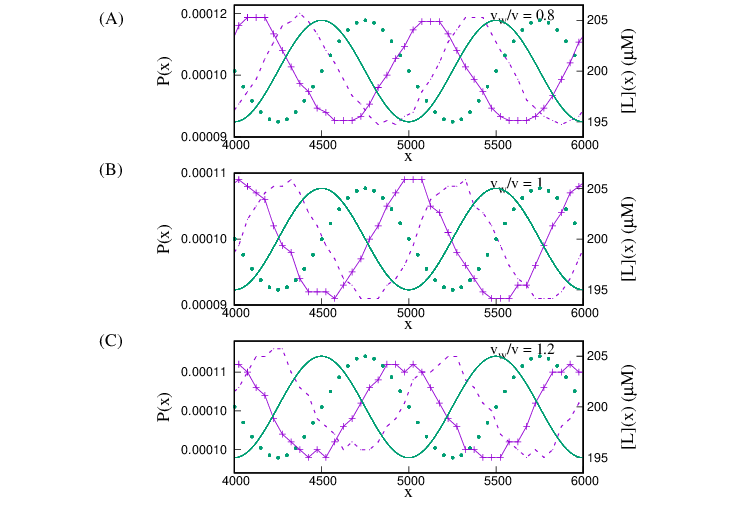}
\caption{Relative position of $P(x,t)$ (violet) and $[L](x,t)$ (green) have been shown for three different values of $v_w$ for moderate wavelength. The solid lines show these two waves at $t=3T/4$ and the dotted lines show them at a later time $t=T$. As $v_w$ increases, the peak of $P(x,t)$ is further away from the peak of $[L](x,t)$. All simulation parameters are as in Fig. \ref{fig:v100}.}
\label{fig:pxcxlg}
\end{figure}

\begin{figure}
\includegraphics[scale=1]{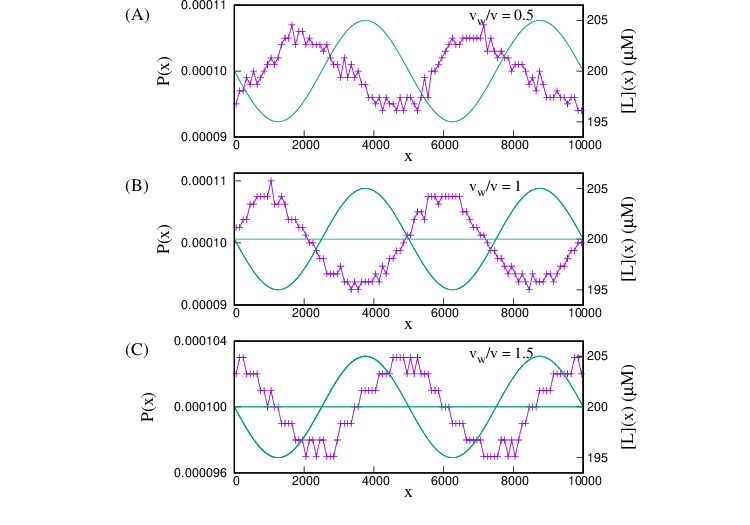}
\caption{Relative position of $P(x,t)$ (violet) and $[L](x,t)$ (green) have been shown for three different values of $v_w$ in large wavelength case. As $v_w$ increases, the phase difference between the two waves also increases. This is consistent with our data in Fig. \ref{fig:lag}. All other simulation parameters are as in Fig. \ref{fig:v500}.}
\label{fig:pxcxlgL}
\end{figure}

Our data in Fig. \ref{fig:lag} show that  $\Delta \phi$ increases with wave speed. For very small $v_w$, we find $\Delta \phi$ is close to zero, which indicates there is hardly any phase lag between cell density wave and the attractant wave for small propagation speed. However, the time-averaged drift velocity of a single cell is still negative in this range, as shown in Fig. \ref{fig:v500}. When the wave speed is roughly equal to the run speed of the cell, $\Delta \phi \sim \pi$, which means peak of $P(x,t)$ is found near the minimum of $[L](x,t)$ (Fig. \ref{fig:pxcxlg}B and Fig. \ref{fig:pxcxlgL}B). For wave speed twice as much as the run speed, $\Delta \phi$ is close to $2 \pi$, {\sl i.e.} the cell density lags behind the attractant wave by one complete time period. A similar effect was observed in an earlier experiment \cite{li2017barrier} where traveling wave attractant profile was set up inside a microfluidic channel and the maximum of cell density was found to move backward for larger wave speed, and finally when the wave speed is significantly large, the large density patch is observed at the next peak of the attractant wave. In \cite{zhu2012frequency} a time-periodic (but not spatially periodic) attractant profile was created inside a microfluidic channel and it was found that the bacterial population at a given location also oscillates with time with the same periodicity, and with a phase lag. It was shown that the phase lag decreases with time period, which is consistent with our data in Fig. \ref{fig:lag}. 

\section{Runs originating from rising and falling parts of the wave} \label{app:wake}

\begin{figure}
\includegraphics[scale=1.5]{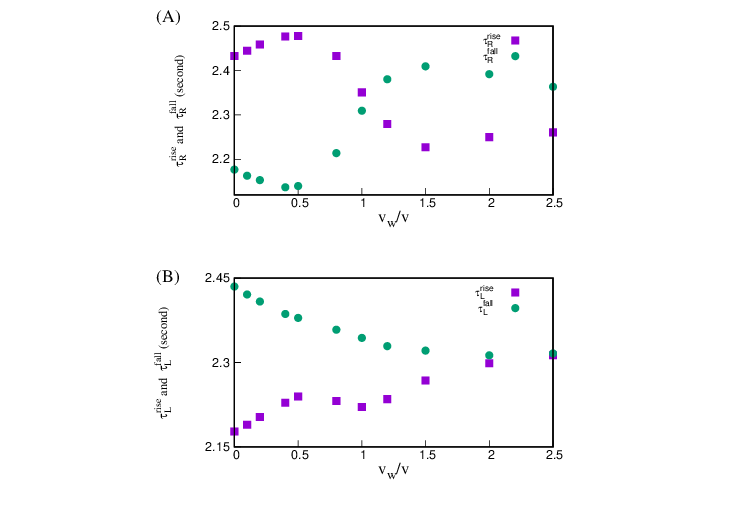}
\caption{Average duration of rightward (A) and leftward (B) runs originating from the rising and falling phases of the attractant wave, {\sl vs} the wave speed. The maximum error-bar in these data is $0.01s$. These data are for moderate $\lambda$ and all simulation parameters are as in Fig. \ref{fig:v100}.}
\label{fig:wke100}
\end{figure}

\begin{figure}
\includegraphics[scale=1.5]{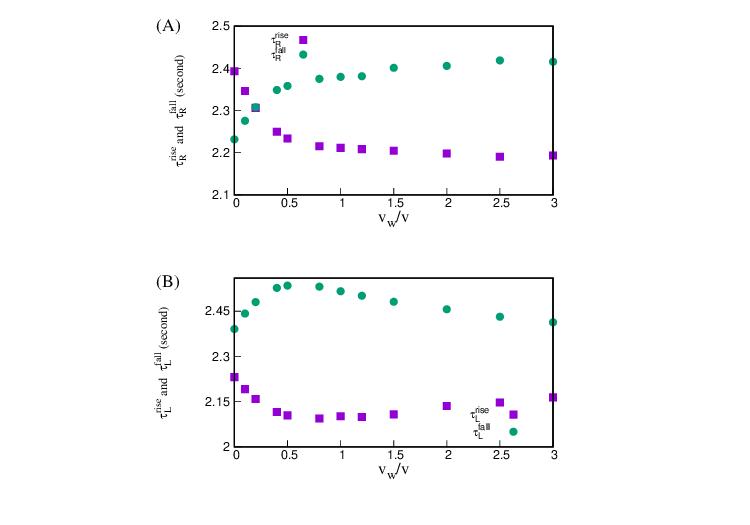}
\caption{Average duration of rightward (A) and leftward (B) runs originating from the rising and falling phases of the attractant wave, {\sl vs} the wave speed. The maximum error-bar in these data is $0.01s$. These data are for large $\lambda$ and all simulation parameters are as in Fig. \ref{fig:v500}.}
\label{fig:wke500}
\end{figure}

In Sec. \ref{sec:hilo} we discussed runs starting from regions of low and high attractant levels. In this appendix we look at runs which are starting from rising and falling phases of the attractant wave. A rightward run starting from the rising part of the wave, initially experiences increasing (decreasing) attractant concentration for $v_w$ less (greater) than $v$. Moreover, as $v_w$ increases, the effective gradient seen by the rightmover becomes weaker for $v_w <v$ and steeper for $v_w > v$. This means $\tau_R^{rise}$, the mean rightward run duration starting from the rising parts of the wave decreases with $v_w$. Our data in Fig. \ref{fig:wke100}A confirms this trend except for very small and very large $v_w$, when opposite trend is observed. When $v_w \ll v$ or $v_w \gg v$, then the relative velocity between the wave and rightmovers is quite large and for moderate $\lambda$ values, the right-movers during a run often cross the maximum (or minimum) of the traveling sine wave. When this happens, the right-movers experience decreasing (increasing) attractant level. This effect is responsible for the trend reversal.  For leftward runs the effect is even stronger. The relative velocity $-(v+v_w)$ is quite large for all $v_w$ values and most leftward runs experience a change in the sign of the attractant gradient. For example, the leftward runs starting from the rising part of the wave, initially encounter decreasing attractant levels. But they quickly cross over the wave minimum and now find attractant levels increasing along their path. As $v_w$ increases, these runs experience steeper gradient and become longer. Our data in Fig. \ref{fig:wke100}B show this. However, for very large $\lambda$ this crossing over does not happen as frequently and in most cases a run starting up the gradient remains so until it ends. Our data in Fig. \ref{fig:wke500} are consistent with this.

\section{$a_0$ at different attractant zones} \label{app:a0zone}

\begin{figure}
\includegraphics[scale=1]{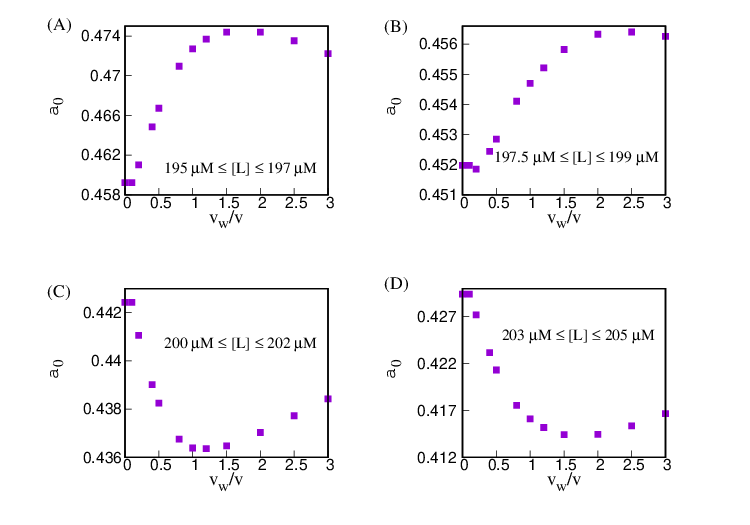}
\caption{Initial activity ($a_0$) of runs starting from different zones of attractant concentration as function of wave speed for large $\lambda$. The zones have been chosen such that the entire concentration range is split in roughly four equal parts, as shown in the legends. The maximum error-bar in these data is $10^{-5}$. All other simulation parameters are as in Fig. \ref{fig:v500}.}
\label{fig:acwhol}
\end{figure}

\begin{figure}
\includegraphics[scale=1]{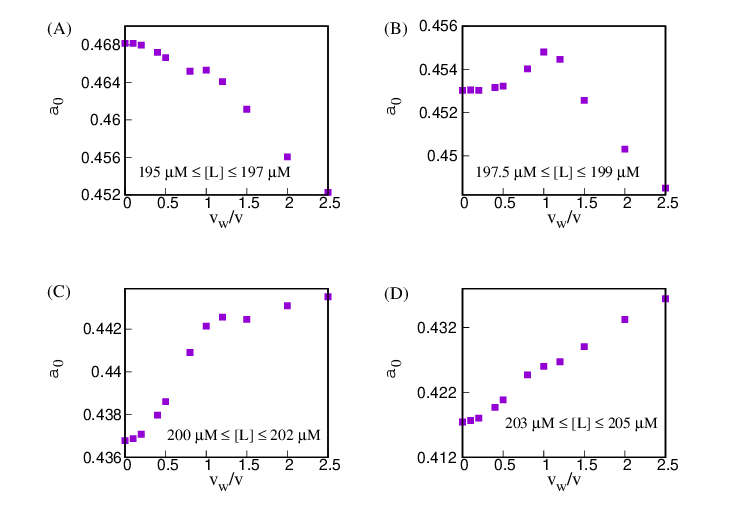}
\caption{Initial activity ($a_0$) of runs starting from different zones of attractant concentration as function of wave speed for moderate wavelength. Zones are defined in each panel, which split the attractant concentration range in four roughly equal parts. The maximum error-bar in these data is $10^{-5}$. All other simulation parameters are as in Fig. \ref{fig:v100}.}
\label{fig:acwhol100}
\end{figure}

In Fig. \ref{fig:a0tr} we have shown the variation of initial activity (measured at the beginning of a run) for runs starting from regions with local attractant concentration lower than the average $[L]_0$. In this appendix we show initial activity for various concentration ranges. From Eq. \ref{eq:cxt} and our choice of parameters, $[L](x,t)$ varies between $195$ and $205$ $\mu M$. We break it up into $4$ smaller ranges and measure $a_0$ for runs starting from each concentration range, as a function of $v_w /v$. Fig. \ref{fig:acwhol} shows data for large wavelength and we find that $a_0$ correctly captures the variation of mean run lengths in Fig. \ref{fig:minmax500}. Figs. \ref{fig:acwhol}A and \ref{fig:acwhol}B consider runs starting with $[L](x,t)$ values in the range $[195,197]$ $\mu M$ and  $[197.5,199]$ $\mu M$, respectively. Both these ranges are below the average value $200$ $\mu M$. We find similar behavior of $a_0$ in both these cases, which is again consistent with $\tau_R^{low}$ and $\tau_L^{low}$ shown in Fig. \ref{fig:minmax500}. Similarly, Figs. \ref{fig:acwhol}C and \ref{fig:acwhol}D consider attractant concentration in the range $[200,202]$ $\mu M$ and  $[203,205]$ $\mu M$. The variation of $a_0$ in these panels is consistent with $\tau_R^{high}$ and $\tau_L^{high}$. On the other hand, Fig. \ref{fig:acwhol100} shows data for moderate wavelength where trends shown by $a_0$ do not always reflect the behavior of mean run lengths in Fig. \ref{fig:minmax100}. In this case, the activity shows larger variation during a run and the initial activity does not completely control the run duration.

%


\begin{thebibliography}{58}%
\makeatletter
\providecommand \@ifxundefined [1]{%
 \@ifx{#1\undefined}
}%
\providecommand \@ifnum [1]{%
 \ifnum #1\expandafter \@firstoftwo
 \else \expandafter \@secondoftwo
 \fi
}%
\providecommand \@ifx [1]{%
 \ifx #1\expandafter \@firstoftwo
 \else \expandafter \@secondoftwo
 \fi
}%
\providecommand \natexlab [1]{#1}%
\providecommand \enquote  [1]{``#1''}%
\providecommand \bibnamefont  [1]{#1}%
\providecommand \bibfnamefont [1]{#1}%
\providecommand \citenamefont [1]{#1}%
\providecommand \href@noop [0]{\@secondoftwo}%
\providecommand \href [0]{\begingroup \@sanitize@url \@href}%
\providecommand \@href[1]{\@@startlink{#1}\@@href}%
\providecommand \@@href[1]{\endgroup#1\@@endlink}%
\providecommand \@sanitize@url [0]{\catcode `\\12\catcode `\$12\catcode
  `\&12\catcode `\#12\catcode `\^12\catcode `\_12\catcode `\%12\relax}%
\providecommand \@@startlink[1]{}%
\providecommand \@@endlink[0]{}%
\providecommand \url  [0]{\begingroup\@sanitize@url \@url }%
\providecommand \@url [1]{\endgroup\@href {#1}{\urlprefix }}%
\providecommand \urlprefix  [0]{URL }%
\providecommand \Eprint [0]{\href }%
\providecommand \doibase [0]{https://doi.org/}%
\providecommand \selectlanguage [0]{\@gobble}%
\providecommand \bibinfo  [0]{\@secondoftwo}%
\providecommand \bibfield  [0]{\@secondoftwo}%
\providecommand \translation [1]{[#1]}%
\providecommand \BibitemOpen [0]{}%
\providecommand \bibitemStop [0]{}%
\providecommand \bibitemNoStop [0]{.\EOS\space}%
\providecommand \EOS [0]{\spacefactor3000\relax}%
\providecommand \BibitemShut  [1]{\csname bibitem#1\endcsname}%
\let\auto@bib@innerbib\@empty
\bibitem [{\citenamefont {Eisenbach}(2004)}]{eisenbach}%
  \BibitemOpen
  \bibfield  {author} {\bibinfo {author} {\bibfnamefont {M.}~\bibnamefont
  {Eisenbach}},\ }\href {https://doi.org/10.1142/p303} {\emph {\bibinfo {title}
  {Chemotaxis}}}\ (\bibinfo  {publisher} {Imperial College Press, Worls
  Scientific Publishing Co.},\ \bibinfo {year} {2004})\BibitemShut {NoStop}%
\bibitem [{\citenamefont {Daniels}\ \emph {et~al.}(2004)\citenamefont
  {Daniels}, \citenamefont {Vanderleyden},\ and\ \citenamefont
  {Michiels}}]{daniels2004quorum}%
  \BibitemOpen
  \bibfield  {author} {\bibinfo {author} {\bibfnamefont {R.}~\bibnamefont
  {Daniels}}, \bibinfo {author} {\bibfnamefont {J.}~\bibnamefont
  {Vanderleyden}},\ and\ \bibinfo {author} {\bibfnamefont {J.}~\bibnamefont
  {Michiels}},\ }\bibfield  {title} {\bibinfo {title} {Quorum sensing and
  swarming migration in bacteria},\ }\href@noop {} {\bibfield  {journal}
  {\bibinfo  {journal} {FEMS microbiology reviews}\ }\textbf {\bibinfo {volume}
  {28}},\ \bibinfo {pages} {261} (\bibinfo {year} {2004})}\BibitemShut
  {NoStop}%
\bibitem [{\citenamefont {Dubravcic}\ \emph {et~al.}(2014)\citenamefont
  {Dubravcic}, \citenamefont {Van~Baalen},\ and\ \citenamefont
  {Nizak}}]{dubravcic2014evolutionarily}%
  \BibitemOpen
  \bibfield  {author} {\bibinfo {author} {\bibfnamefont {D.}~\bibnamefont
  {Dubravcic}}, \bibinfo {author} {\bibfnamefont {M.}~\bibnamefont
  {Van~Baalen}},\ and\ \bibinfo {author} {\bibfnamefont {C.}~\bibnamefont
  {Nizak}},\ }\bibfield  {title} {\bibinfo {title} {An evolutionarily
  significant unicellular strategy in response to starvation in dictyostelium
  social amoebae},\ }\href@noop {} {\bibfield  {journal} {\bibinfo  {journal}
  {F1000Research}\ }\textbf {\bibinfo {volume} {3}} (\bibinfo {year}
  {2014})}\BibitemShut {NoStop}%
\bibitem [{\citenamefont {Cai}\ \emph {et~al.}(2012)\citenamefont {Cai},
  \citenamefont {Huang}, \citenamefont {Devreotes},\ and\ \citenamefont
  {Iijima}}]{cai2012analysis}%
  \BibitemOpen
  \bibfield  {author} {\bibinfo {author} {\bibfnamefont {H.}~\bibnamefont
  {Cai}}, \bibinfo {author} {\bibfnamefont {C.-H.}\ \bibnamefont {Huang}},
  \bibinfo {author} {\bibfnamefont {P.~N.}\ \bibnamefont {Devreotes}},\ and\
  \bibinfo {author} {\bibfnamefont {M.}~\bibnamefont {Iijima}},\ }\bibfield
  {title} {\bibinfo {title} {Analysis of chemotaxis in dictyostelium},\
  }\href@noop {} {\bibfield  {journal} {\bibinfo  {journal} {Integrin and Cell
  Adhesion Molecules: Methods and Protocols}\ ,\ \bibinfo {pages} {451}}
  (\bibinfo {year} {2012})}\BibitemShut {NoStop}%
\bibitem [{\citenamefont {Goldstein}(1996)}]{goldstein1996traveling}%
  \BibitemOpen
  \bibfield  {author} {\bibinfo {author} {\bibfnamefont {R.~E.}\ \bibnamefont
  {Goldstein}},\ }\bibfield  {title} {\bibinfo {title} {Traveling-wave
  chemotaxis},\ }\href@noop {} {\bibfield  {journal} {\bibinfo  {journal}
  {Physical review letters}\ }\textbf {\bibinfo {volume} {77}},\ \bibinfo
  {pages} {775} (\bibinfo {year} {1996})}\BibitemShut {NoStop}%
\bibitem [{\citenamefont {Afonso}\ \emph {et~al.}(2012)\citenamefont {Afonso},
  \citenamefont {Janka-Junttila}, \citenamefont {Lee}, \citenamefont {McCann},
  \citenamefont {Oliver}, \citenamefont {Aamer}, \citenamefont {Losert},
  \citenamefont {Cicerone},\ and\ \citenamefont {Parent}}]{afonso2012ltb4}%
  \BibitemOpen
  \bibfield  {author} {\bibinfo {author} {\bibfnamefont {P.~V.}\ \bibnamefont
  {Afonso}}, \bibinfo {author} {\bibfnamefont {M.}~\bibnamefont
  {Janka-Junttila}}, \bibinfo {author} {\bibfnamefont {Y.~J.}\ \bibnamefont
  {Lee}}, \bibinfo {author} {\bibfnamefont {C.~P.}\ \bibnamefont {McCann}},
  \bibinfo {author} {\bibfnamefont {C.~M.}\ \bibnamefont {Oliver}}, \bibinfo
  {author} {\bibfnamefont {K.~A.}\ \bibnamefont {Aamer}}, \bibinfo {author}
  {\bibfnamefont {W.}~\bibnamefont {Losert}}, \bibinfo {author} {\bibfnamefont
  {M.~T.}\ \bibnamefont {Cicerone}},\ and\ \bibinfo {author} {\bibfnamefont
  {C.~A.}\ \bibnamefont {Parent}},\ }\bibfield  {title} {\bibinfo {title} {Ltb4
  is a signal-relay molecule during neutrophil chemotaxis},\ }\href@noop {}
  {\bibfield  {journal} {\bibinfo  {journal} {Developmental cell}\ }\textbf
  {\bibinfo {volume} {22}},\ \bibinfo {pages} {1079} (\bibinfo {year}
  {2012})}\BibitemShut {NoStop}%
\bibitem [{\citenamefont {Ishida}\ \emph {et~al.}(2025)\citenamefont {Ishida},
  \citenamefont {Uwamichi}, \citenamefont {Nakajima},\ and\ \citenamefont
  {Sawai}}]{ishida2025traveling}%
  \BibitemOpen
  \bibfield  {author} {\bibinfo {author} {\bibfnamefont {M.}~\bibnamefont
  {Ishida}}, \bibinfo {author} {\bibfnamefont {M.}~\bibnamefont {Uwamichi}},
  \bibinfo {author} {\bibfnamefont {A.}~\bibnamefont {Nakajima}},\ and\
  \bibinfo {author} {\bibfnamefont {S.}~\bibnamefont {Sawai}},\ }\bibfield
  {title} {\bibinfo {title} {Traveling wave chemotaxis of neutrophil-like hl-60
  cells},\ }\href@noop {} {\bibfield  {journal} {\bibinfo  {journal} {Molecular
  Biology of the Cell}\ ,\ \bibinfo {pages} {mbc}} (\bibinfo {year}
  {2025})}\BibitemShut {NoStop}%
\bibitem [{\citenamefont {Tweedy}\ \emph {et~al.}(2016)\citenamefont {Tweedy},
  \citenamefont {Knecht}, \citenamefont {Mackay},\ and\ \citenamefont
  {Insall}}]{tweedy2016self}%
  \BibitemOpen
  \bibfield  {author} {\bibinfo {author} {\bibfnamefont {L.}~\bibnamefont
  {Tweedy}}, \bibinfo {author} {\bibfnamefont {D.~A.}\ \bibnamefont {Knecht}},
  \bibinfo {author} {\bibfnamefont {G.~M.}\ \bibnamefont {Mackay}},\ and\
  \bibinfo {author} {\bibfnamefont {R.~H.}\ \bibnamefont {Insall}},\ }\bibfield
   {title} {\bibinfo {title} {Self-generated chemoattractant gradients:
  attractant depletion extends the range and robustness of chemotaxis},\
  }\href@noop {} {\bibfield  {journal} {\bibinfo  {journal} {PLoS biology}\
  }\textbf {\bibinfo {volume} {14}},\ \bibinfo {pages} {e1002404} (\bibinfo
  {year} {2016})}\BibitemShut {NoStop}%
\bibitem [{\citenamefont {Goldbeter}(1997)}]{goldbeter1997modelling}%
  \BibitemOpen
  \bibfield  {author} {\bibinfo {author} {\bibfnamefont {A.}~\bibnamefont
  {Goldbeter}},\ }\bibfield  {title} {\bibinfo {title} {Modelling biochemical
  oscillations and cellular rhythms},\ }\href@noop {} {\bibfield  {journal}
  {\bibinfo  {journal} {Current Science}\ ,\ \bibinfo {pages} {933}} (\bibinfo
  {year} {1997})}\BibitemShut {NoStop}%
\bibitem [{\citenamefont {Berg}(2008)}]{2008cbergoli}%
  \BibitemOpen
  \bibfield  {author} {\bibinfo {author} {\bibfnamefont {H.~C.}\ \bibnamefont
  {Berg}},\ }\href@noop {} {\emph {\bibinfo {title} {E. coli in Motion}}}\
  (\bibinfo  {publisher} {Springer Science \& Business Media},\ \bibinfo {year}
  {2008})\BibitemShut {NoStop}%
\bibitem [{\citenamefont {Tu}(2013)}]{tu2013quantitative}%
  \BibitemOpen
  \bibfield  {author} {\bibinfo {author} {\bibfnamefont {Y.}~\bibnamefont
  {Tu}},\ }\bibfield  {title} {\bibinfo {title} {Quantitative modeling of
  bacterial chemotaxis: signal amplification and accurate adaptation},\
  }\href@noop {} {\bibfield  {journal} {\bibinfo  {journal} {Annual review of
  biophysics}\ }\textbf {\bibinfo {volume} {42}},\ \bibinfo {pages} {337}
  (\bibinfo {year} {2013})}\BibitemShut {NoStop}%
\bibitem [{\citenamefont {Tu}\ \emph {et~al.}(2008)\citenamefont {Tu},
  \citenamefont {Shimizu},\ and\ \citenamefont {Berg}}]{tu2008modeling}%
  \BibitemOpen
  \bibfield  {author} {\bibinfo {author} {\bibfnamefont {Y.}~\bibnamefont
  {Tu}}, \bibinfo {author} {\bibfnamefont {T.~S.}\ \bibnamefont {Shimizu}},\
  and\ \bibinfo {author} {\bibfnamefont {H.~C.}\ \bibnamefont {Berg}},\
  }\bibfield  {title} {\bibinfo {title} {Modeling the chemotactic response of
  escherichia coli to time-varying stimuli},\ }\href@noop {} {\bibfield
  {journal} {\bibinfo  {journal} {Proceedings of the National Academy of
  Sciences}\ }\textbf {\bibinfo {volume} {105}},\ \bibinfo {pages} {14855}
  (\bibinfo {year} {2008})}\BibitemShut {NoStop}%
\bibitem [{\citenamefont {Shimizu}\ \emph {et~al.}(2010)\citenamefont
  {Shimizu}, \citenamefont {Tu},\ and\ \citenamefont
  {Berg}}]{shimizu2010modular}%
  \BibitemOpen
  \bibfield  {author} {\bibinfo {author} {\bibfnamefont {T.~S.}\ \bibnamefont
  {Shimizu}}, \bibinfo {author} {\bibfnamefont {Y.}~\bibnamefont {Tu}},\ and\
  \bibinfo {author} {\bibfnamefont {H.~C.}\ \bibnamefont {Berg}},\ }\bibfield
  {title} {\bibinfo {title} {A modular gradient-sensing network for chemotaxis
  in escherichia coli revealed by responses to time-varying stimuli},\
  }\href@noop {} {\bibfield  {journal} {\bibinfo  {journal} {Molecular systems
  biology}\ }\textbf {\bibinfo {volume} {6}},\ \bibinfo {pages} {382} (\bibinfo
  {year} {2010})}\BibitemShut {NoStop}%
\bibitem [{\citenamefont {Lazova}\ \emph {et~al.}(2011)\citenamefont {Lazova},
  \citenamefont {Ahmed}, \citenamefont {Bellomo}, \citenamefont {Stocker},\
  and\ \citenamefont {Shimizu}}]{lazova2011response}%
  \BibitemOpen
  \bibfield  {author} {\bibinfo {author} {\bibfnamefont {M.~D.}\ \bibnamefont
  {Lazova}}, \bibinfo {author} {\bibfnamefont {T.}~\bibnamefont {Ahmed}},
  \bibinfo {author} {\bibfnamefont {D.}~\bibnamefont {Bellomo}}, \bibinfo
  {author} {\bibfnamefont {R.}~\bibnamefont {Stocker}},\ and\ \bibinfo {author}
  {\bibfnamefont {T.~S.}\ \bibnamefont {Shimizu}},\ }\bibfield  {title}
  {\bibinfo {title} {Response rescaling in bacterial chemotaxis},\ }\href@noop
  {} {\bibfield  {journal} {\bibinfo  {journal} {Proceedings of the National
  Academy of Sciences}\ }\textbf {\bibinfo {volume} {108}},\ \bibinfo {pages}
  {13870} (\bibinfo {year} {2011})}\BibitemShut {NoStop}%
\bibitem [{\citenamefont {Zhu}\ \emph {et~al.}(2012)\citenamefont {Zhu},
  \citenamefont {Si}, \citenamefont {Deng}, \citenamefont {Ouyang},
  \citenamefont {Wu}, \citenamefont {He}, \citenamefont {Jiang}, \citenamefont
  {Luo},\ and\ \citenamefont {Tu}}]{zhu2012frequency}%
  \BibitemOpen
  \bibfield  {author} {\bibinfo {author} {\bibfnamefont {X.}~\bibnamefont
  {Zhu}}, \bibinfo {author} {\bibfnamefont {G.}~\bibnamefont {Si}}, \bibinfo
  {author} {\bibfnamefont {N.}~\bibnamefont {Deng}}, \bibinfo {author}
  {\bibfnamefont {Q.}~\bibnamefont {Ouyang}}, \bibinfo {author} {\bibfnamefont
  {T.}~\bibnamefont {Wu}}, \bibinfo {author} {\bibfnamefont {Z.}~\bibnamefont
  {He}}, \bibinfo {author} {\bibfnamefont {L.}~\bibnamefont {Jiang}}, \bibinfo
  {author} {\bibfnamefont {C.}~\bibnamefont {Luo}},\ and\ \bibinfo {author}
  {\bibfnamefont {Y.}~\bibnamefont {Tu}},\ }\bibfield  {title} {\bibinfo
  {title} {Frequency-dependent escherichia coli chemotaxis behavior},\
  }\href@noop {} {\bibfield  {journal} {\bibinfo  {journal} {Physical review
  letters}\ }\textbf {\bibinfo {volume} {108}},\ \bibinfo {pages} {128101}
  (\bibinfo {year} {2012})}\BibitemShut {NoStop}%
\bibitem [{\citenamefont {Jiang}\ \emph {et~al.}(2010)\citenamefont {Jiang},
  \citenamefont {Ouyang},\ and\ \citenamefont {Tu}}]{jiang2010quantitative}%
  \BibitemOpen
  \bibfield  {author} {\bibinfo {author} {\bibfnamefont {L.}~\bibnamefont
  {Jiang}}, \bibinfo {author} {\bibfnamefont {Q.}~\bibnamefont {Ouyang}},\ and\
  \bibinfo {author} {\bibfnamefont {Y.}~\bibnamefont {Tu}},\ }\bibfield
  {title} {\bibinfo {title} {Quantitative modeling of escherichia coli
  chemotactic motion in environments varying in space and time},\ }\href@noop
  {} {\bibfield  {journal} {\bibinfo  {journal} {PLoS Comput Biol}\ }\textbf
  {\bibinfo {volume} {6}},\ \bibinfo {pages} {e1000735} (\bibinfo {year}
  {2010})}\BibitemShut {NoStop}%
\bibitem [{\citenamefont {Li}\ \emph {et~al.}(2017)\citenamefont {Li},
  \citenamefont {Cai}, \citenamefont {Zhang}, \citenamefont {Si}, \citenamefont
  {Ouyang}, \citenamefont {Luo},\ and\ \citenamefont {Tu}}]{li2017barrier}%
  \BibitemOpen
  \bibfield  {author} {\bibinfo {author} {\bibfnamefont {Z.}~\bibnamefont
  {Li}}, \bibinfo {author} {\bibfnamefont {Q.}~\bibnamefont {Cai}}, \bibinfo
  {author} {\bibfnamefont {X.}~\bibnamefont {Zhang}}, \bibinfo {author}
  {\bibfnamefont {G.}~\bibnamefont {Si}}, \bibinfo {author} {\bibfnamefont
  {Q.}~\bibnamefont {Ouyang}}, \bibinfo {author} {\bibfnamefont
  {C.}~\bibnamefont {Luo}},\ and\ \bibinfo {author} {\bibfnamefont
  {Y.}~\bibnamefont {Tu}},\ }\bibfield  {title} {\bibinfo {title} {Barrier
  crossing in escherichia coli chemotaxis},\ }\href@noop {} {\bibfield
  {journal} {\bibinfo  {journal} {Physical review letters}\ }\textbf {\bibinfo
  {volume} {118}},\ \bibinfo {pages} {098101} (\bibinfo {year}
  {2017})}\BibitemShut {NoStop}%
\bibitem [{\citenamefont {Mandal}\ and\ \citenamefont
  {Chatterjee}(2021)}]{shobhan}%
  \BibitemOpen
  \bibfield  {author} {\bibinfo {author} {\bibfnamefont {S.~D.}\ \bibnamefont
  {Mandal}}\ and\ \bibinfo {author} {\bibfnamefont {S.}~\bibnamefont
  {Chatterjee}},\ }\bibfield  {title} {\bibinfo {title} {Effect of receptor
  clustering on chemotactic performance of e. coli: Sensing versus
  adaptation},\ }\href {https://doi.org/10.1103/PhysRevE.103.L030401}
  {\bibfield  {journal} {\bibinfo  {journal} {Phys. Rev. E}\ }\textbf {\bibinfo
  {volume} {103}},\ \bibinfo {pages} {L030401} (\bibinfo {year}
  {2021})}\BibitemShut {NoStop}%
\bibitem [{\citenamefont {Mandal}\ and\ \citenamefont
  {Chatterjee}(2022{\natexlab{a}})}]{mandal2022effectre}%
  \BibitemOpen
  \bibfield  {author} {\bibinfo {author} {\bibfnamefont {S.~D.}\ \bibnamefont
  {Mandal}}\ and\ \bibinfo {author} {\bibfnamefont {S.}~\bibnamefont
  {Chatterjee}},\ }\bibfield  {title} {\bibinfo {title} {Effect of switching
  time scale of receptor activity on chemotactic performance of escherichia
  coli},\ }\href@noop {} {\bibfield  {journal} {\bibinfo  {journal} {Indian
  Journal of Physics}\ }\textbf {\bibinfo {volume} {96}},\ \bibinfo {pages}
  {2619} (\bibinfo {year} {2022}{\natexlab{a}})}\BibitemShut {NoStop}%
\bibitem [{\citenamefont {Mandal}\ and\ \citenamefont
  {Chatterjee}(2022{\natexlab{b}})}]{mandal2022effect}%
  \BibitemOpen
  \bibfield  {author} {\bibinfo {author} {\bibfnamefont {S.~D.}\ \bibnamefont
  {Mandal}}\ and\ \bibinfo {author} {\bibfnamefont {S.}~\bibnamefont
  {Chatterjee}},\ }\bibfield  {title} {\bibinfo {title} {Effect of receptor
  cooperativity on methylation dynamics in bacterial chemotaxis with weak and
  strong gradient},\ }\href@noop {} {\bibfield  {journal} {\bibinfo  {journal}
  {Physical Review E}\ }\textbf {\bibinfo {volume} {105}},\ \bibinfo {pages}
  {014411} (\bibinfo {year} {2022}{\natexlab{b}})}\BibitemShut {NoStop}%
\bibitem [{\citenamefont {Pontius}\ \emph {et~al.}(2013)\citenamefont
  {Pontius}, \citenamefont {Sneddon},\ and\ \citenamefont
  {Emonet}}]{pontius2013adaptation}%
  \BibitemOpen
  \bibfield  {author} {\bibinfo {author} {\bibfnamefont {W.}~\bibnamefont
  {Pontius}}, \bibinfo {author} {\bibfnamefont {M.~W.}\ \bibnamefont
  {Sneddon}},\ and\ \bibinfo {author} {\bibfnamefont {T.}~\bibnamefont
  {Emonet}},\ }\bibfield  {title} {\bibinfo {title} {Adaptation dynamics in
  densely clustered chemoreceptors},\ }\href@noop {} {\bibfield  {journal}
  {\bibinfo  {journal} {PLoS computational biology}\ }\textbf {\bibinfo
  {volume} {9}} (\bibinfo {year} {2013})}\BibitemShut {NoStop}%
\bibitem [{\citenamefont {Briegel}\ \emph {et~al.}(2012)\citenamefont
  {Briegel}, \citenamefont {Li}, \citenamefont {Bilwes}, \citenamefont
  {Hughes}, \citenamefont {Jensen},\ and\ \citenamefont
  {Crane}}]{briegel2012bacterial}%
  \BibitemOpen
  \bibfield  {author} {\bibinfo {author} {\bibfnamefont {A.}~\bibnamefont
  {Briegel}}, \bibinfo {author} {\bibfnamefont {X.}~\bibnamefont {Li}},
  \bibinfo {author} {\bibfnamefont {A.~M.}\ \bibnamefont {Bilwes}}, \bibinfo
  {author} {\bibfnamefont {K.~T.}\ \bibnamefont {Hughes}}, \bibinfo {author}
  {\bibfnamefont {G.~J.}\ \bibnamefont {Jensen}},\ and\ \bibinfo {author}
  {\bibfnamefont {B.~R.}\ \bibnamefont {Crane}},\ }\bibfield  {title} {\bibinfo
  {title} {Bacterial chemoreceptor arrays are hexagonally packed trimers of
  receptor dimers networked by rings of kinase and coupling proteins},\
  }\href@noop {} {\bibfield  {journal} {\bibinfo  {journal} {Proceedings of the
  National Academy of Sciences}\ }\textbf {\bibinfo {volume} {109}},\ \bibinfo
  {pages} {3766} (\bibinfo {year} {2012})}\BibitemShut {NoStop}%
\bibitem [{\citenamefont {Liu}\ \emph {et~al.}(2012)\citenamefont {Liu},
  \citenamefont {Hu}, \citenamefont {Morado}, \citenamefont {Jani},
  \citenamefont {Manson},\ and\ \citenamefont {Margolin}}]{liu2012molecular}%
  \BibitemOpen
  \bibfield  {author} {\bibinfo {author} {\bibfnamefont {J.}~\bibnamefont
  {Liu}}, \bibinfo {author} {\bibfnamefont {B.}~\bibnamefont {Hu}}, \bibinfo
  {author} {\bibfnamefont {D.~R.}\ \bibnamefont {Morado}}, \bibinfo {author}
  {\bibfnamefont {S.}~\bibnamefont {Jani}}, \bibinfo {author} {\bibfnamefont
  {M.~D.}\ \bibnamefont {Manson}},\ and\ \bibinfo {author} {\bibfnamefont
  {W.}~\bibnamefont {Margolin}},\ }\bibfield  {title} {\bibinfo {title}
  {Molecular architecture of chemoreceptor arrays revealed by cryoelectron
  tomography of escherichia coli minicells},\ }\href@noop {} {\bibfield
  {journal} {\bibinfo  {journal} {Proceedings of National Academy of Sciences}\
  }\textbf {\bibinfo {volume} {109}},\ \bibinfo {pages} {E1481} (\bibinfo
  {year} {2012})}\BibitemShut {NoStop}%
\bibitem [{\citenamefont {Mello}\ and\ \citenamefont
  {Tu}(2005)}]{mello2005allosteric}%
  \BibitemOpen
  \bibfield  {author} {\bibinfo {author} {\bibfnamefont {B.~A.}\ \bibnamefont
  {Mello}}\ and\ \bibinfo {author} {\bibfnamefont {Y.}~\bibnamefont {Tu}},\
  }\bibfield  {title} {\bibinfo {title} {An allosteric model for heterogeneous
  receptor complexes: understanding bacterial chemotaxis responses to multiple
  stimuli},\ }\href@noop {} {\bibfield  {journal} {\bibinfo  {journal}
  {Proceedings of the National Academy of Sciences}\ }\textbf {\bibinfo
  {volume} {102}},\ \bibinfo {pages} {17354} (\bibinfo {year}
  {2005})}\BibitemShut {NoStop}%
\bibitem [{\citenamefont {Monod}\ \emph {et~al.}(1965)\citenamefont {Monod},
  \citenamefont {Wyman},\ and\ \citenamefont {Changeux}}]{monod1965nature}%
  \BibitemOpen
  \bibfield  {author} {\bibinfo {author} {\bibfnamefont {J.}~\bibnamefont
  {Monod}}, \bibinfo {author} {\bibfnamefont {J.}~\bibnamefont {Wyman}},\ and\
  \bibinfo {author} {\bibfnamefont {J.-P.}\ \bibnamefont {Changeux}},\
  }\bibfield  {title} {\bibinfo {title} {On the nature of allosteric
  transitions: a plausible model},\ }\href@noop {} {\bibfield  {journal}
  {\bibinfo  {journal} {J Mol Biol}\ }\textbf {\bibinfo {volume} {12}},\
  \bibinfo {pages} {88} (\bibinfo {year} {1965})}\BibitemShut {NoStop}%
\bibitem [{\citenamefont {Keymer}\ \emph {et~al.}(2006)\citenamefont {Keymer},
  \citenamefont {Endres}, \citenamefont {Skoge}, \citenamefont {Meir},\ and\
  \citenamefont {Wingreen}}]{keymer2006chemosensing}%
  \BibitemOpen
  \bibfield  {author} {\bibinfo {author} {\bibfnamefont {J.~E.}\ \bibnamefont
  {Keymer}}, \bibinfo {author} {\bibfnamefont {R.~G.}\ \bibnamefont {Endres}},
  \bibinfo {author} {\bibfnamefont {M.}~\bibnamefont {Skoge}}, \bibinfo
  {author} {\bibfnamefont {Y.}~\bibnamefont {Meir}},\ and\ \bibinfo {author}
  {\bibfnamefont {N.~S.}\ \bibnamefont {Wingreen}},\ }\bibfield  {title}
  {\bibinfo {title} {Chemosensing in escherichia coli: two regimes of two-state
  receptors},\ }\href@noop {} {\bibfield  {journal} {\bibinfo  {journal}
  {Proceedings of the National Academy of Sciences}\ }\textbf {\bibinfo
  {volume} {103}},\ \bibinfo {pages} {1786} (\bibinfo {year}
  {2006})}\BibitemShut {NoStop}%
\bibitem [{\citenamefont {Dufour}\ \emph {et~al.}(2014)\citenamefont {Dufour},
  \citenamefont {Fu}, \citenamefont {Hernandez-Nunez},\ and\ \citenamefont
  {Emonet}}]{dufour2014limits}%
  \BibitemOpen
  \bibfield  {author} {\bibinfo {author} {\bibfnamefont {Y.~S.}\ \bibnamefont
  {Dufour}}, \bibinfo {author} {\bibfnamefont {X.}~\bibnamefont {Fu}}, \bibinfo
  {author} {\bibfnamefont {L.}~\bibnamefont {Hernandez-Nunez}},\ and\ \bibinfo
  {author} {\bibfnamefont {T.}~\bibnamefont {Emonet}},\ }\bibfield  {title}
  {\bibinfo {title} {Limits of feedback control in bacterial chemotaxis},\
  }\href@noop {} {\bibfield  {journal} {\bibinfo  {journal} {PLoS Comput Biol}\
  }\textbf {\bibinfo {volume} {10}},\ \bibinfo {pages} {e1003694} (\bibinfo
  {year} {2014})}\BibitemShut {NoStop}%
\bibitem [{\citenamefont {Frankel}\ \emph {et~al.}(2014)\citenamefont
  {Frankel}, \citenamefont {Pontius}, \citenamefont {Dufour}, \citenamefont
  {Long}, \citenamefont {Hernandez-Nunez},\ and\ \citenamefont
  {Emonet}}]{frankel2014adaptability}%
  \BibitemOpen
  \bibfield  {author} {\bibinfo {author} {\bibfnamefont {N.~W.}\ \bibnamefont
  {Frankel}}, \bibinfo {author} {\bibfnamefont {W.}~\bibnamefont {Pontius}},
  \bibinfo {author} {\bibfnamefont {Y.~S.}\ \bibnamefont {Dufour}}, \bibinfo
  {author} {\bibfnamefont {J.}~\bibnamefont {Long}}, \bibinfo {author}
  {\bibfnamefont {L.}~\bibnamefont {Hernandez-Nunez}},\ and\ \bibinfo {author}
  {\bibfnamefont {T.}~\bibnamefont {Emonet}},\ }\bibfield  {title} {\bibinfo
  {title} {Adaptability of non-genetic diversity in bacterial chemotaxis},\
  }\href@noop {} {\bibfield  {journal} {\bibinfo  {journal} {Elife}\ }\textbf
  {\bibinfo {volume} {3}},\ \bibinfo {pages} {e03526} (\bibinfo {year}
  {2014})}\BibitemShut {NoStop}%
\bibitem [{\citenamefont {Long}\ \emph {et~al.}(2017)\citenamefont {Long},
  \citenamefont {Zucker},\ and\ \citenamefont {Emonet}}]{long2017feedback}%
  \BibitemOpen
  \bibfield  {author} {\bibinfo {author} {\bibfnamefont {J.}~\bibnamefont
  {Long}}, \bibinfo {author} {\bibfnamefont {S.~W.}\ \bibnamefont {Zucker}},\
  and\ \bibinfo {author} {\bibfnamefont {T.}~\bibnamefont {Emonet}},\
  }\bibfield  {title} {\bibinfo {title} {Feedback between motion and sensation
  provides nonlinear boost in run-and-tumble navigation},\ }\href@noop {}
  {\bibfield  {journal} {\bibinfo  {journal} {PLoS computational biology}\
  }\textbf {\bibinfo {volume} {13}},\ \bibinfo {pages} {e1005429} (\bibinfo
  {year} {2017})}\BibitemShut {NoStop}%
\bibitem [{\citenamefont {Colin}\ \emph {et~al.}(2017)\citenamefont {Colin},
  \citenamefont {Rosazza}, \citenamefont {Vaknin},\ and\ \citenamefont
  {Sourjik}}]{colin2017multiple}%
  \BibitemOpen
  \bibfield  {author} {\bibinfo {author} {\bibfnamefont {R.}~\bibnamefont
  {Colin}}, \bibinfo {author} {\bibfnamefont {C.}~\bibnamefont {Rosazza}},
  \bibinfo {author} {\bibfnamefont {A.}~\bibnamefont {Vaknin}},\ and\ \bibinfo
  {author} {\bibfnamefont {V.}~\bibnamefont {Sourjik}},\ }\bibfield  {title}
  {\bibinfo {title} {Multiple sources of slow activity fluctuations in a
  bacterial chemosensory network},\ }\href@noop {} {\bibfield  {journal}
  {\bibinfo  {journal} {Elife}\ }\textbf {\bibinfo {volume} {6}},\ \bibinfo
  {pages} {e26796} (\bibinfo {year} {2017})}\BibitemShut {NoStop}%
\bibitem [{\citenamefont {Chatterjee}(2022)}]{chatterjee2022short}%
  \BibitemOpen
  \bibfield  {author} {\bibinfo {author} {\bibfnamefont {S.}~\bibnamefont
  {Chatterjee}},\ }\bibfield  {title} {\bibinfo {title} {Short time extremal
  response to step stimulus for a single cell e. coli},\ }\href@noop {}
  {\bibfield  {journal} {\bibinfo  {journal} {Journal of Statistical Mechanics:
  Theory and Experiment}\ }\textbf {\bibinfo {volume} {2022}},\ \bibinfo
  {pages} {123503} (\bibinfo {year} {2022})}\BibitemShut {NoStop}%
\bibitem [{\citenamefont {Flores}\ \emph {et~al.}(2012)\citenamefont {Flores},
  \citenamefont {Shimizu}, \citenamefont {ten Wolde},\ and\ \citenamefont
  {Tostevin}}]{flores2012signaling}%
  \BibitemOpen
  \bibfield  {author} {\bibinfo {author} {\bibfnamefont {M.}~\bibnamefont
  {Flores}}, \bibinfo {author} {\bibfnamefont {T.~S.}\ \bibnamefont {Shimizu}},
  \bibinfo {author} {\bibfnamefont {P.~R.}\ \bibnamefont {ten Wolde}},\ and\
  \bibinfo {author} {\bibfnamefont {F.}~\bibnamefont {Tostevin}},\ }\bibfield
  {title} {\bibinfo {title} {Signaling noise enhances chemotactic drift of e.
  coli},\ }\href@noop {} {\bibfield  {journal} {\bibinfo  {journal} {Physical
  review letters}\ }\textbf {\bibinfo {volume} {109}},\ \bibinfo {pages}
  {148101} (\bibinfo {year} {2012})}\BibitemShut {NoStop}%
\bibitem [{\citenamefont {Dev}\ and\ \citenamefont
  {Chatterjee}(2018)}]{dev2018optimal}%
  \BibitemOpen
  \bibfield  {author} {\bibinfo {author} {\bibfnamefont {S.}~\bibnamefont
  {Dev}}\ and\ \bibinfo {author} {\bibfnamefont {S.}~\bibnamefont
  {Chatterjee}},\ }\bibfield  {title} {\bibinfo {title} {Optimal methylation
  noise for best chemotactic performance of e. coli},\ }\href@noop {}
  {\bibfield  {journal} {\bibinfo  {journal} {Physical Review E}\ }\textbf
  {\bibinfo {volume} {97}},\ \bibinfo {pages} {032420} (\bibinfo {year}
  {2018})}\BibitemShut {NoStop}%
\bibitem [{\citenamefont {Sneddon}\ \emph {et~al.}(2012)\citenamefont
  {Sneddon}, \citenamefont {Pontius},\ and\ \citenamefont
  {Emonet}}]{sneddon2012stochastic}%
  \BibitemOpen
  \bibfield  {author} {\bibinfo {author} {\bibfnamefont {M.~W.}\ \bibnamefont
  {Sneddon}}, \bibinfo {author} {\bibfnamefont {W.}~\bibnamefont {Pontius}},\
  and\ \bibinfo {author} {\bibfnamefont {T.}~\bibnamefont {Emonet}},\
  }\bibfield  {title} {\bibinfo {title} {Stochastic coordination of multiple
  actuators reduces latency and improves chemotactic response in bacteria},\
  }\href@noop {} {\bibfield  {journal} {\bibinfo  {journal} {Proceedings of the
  National Academy of Sciences}\ }\textbf {\bibinfo {volume} {109}},\ \bibinfo
  {pages} {805} (\bibinfo {year} {2012})}\BibitemShut {NoStop}%
\bibitem [{\citenamefont {Micali}\ \emph {et~al.}(2017)\citenamefont {Micali},
  \citenamefont {Colin}, \citenamefont {Sourjik},\ and\ \citenamefont
  {Endres}}]{micali2017drift}%
  \BibitemOpen
  \bibfield  {author} {\bibinfo {author} {\bibfnamefont {G.}~\bibnamefont
  {Micali}}, \bibinfo {author} {\bibfnamefont {R.}~\bibnamefont {Colin}},
  \bibinfo {author} {\bibfnamefont {V.}~\bibnamefont {Sourjik}},\ and\ \bibinfo
  {author} {\bibfnamefont {R.~G.}\ \bibnamefont {Endres}},\ }\bibfield  {title}
  {\bibinfo {title} {Drift and behavior of e. coli cells},\ }\href@noop {}
  {\bibfield  {journal} {\bibinfo  {journal} {Biophysical journal}\ }\textbf
  {\bibinfo {volume} {113}},\ \bibinfo {pages} {2321} (\bibinfo {year}
  {2017})}\BibitemShut {NoStop}%
\bibitem [{\citenamefont {Berg}\ and\ \citenamefont
  {Brown}(1972)}]{berg1972chemotaxis}%
  \BibitemOpen
  \bibfield  {author} {\bibinfo {author} {\bibfnamefont {H.~C.}\ \bibnamefont
  {Berg}}\ and\ \bibinfo {author} {\bibfnamefont {D.~A.}\ \bibnamefont
  {Brown}},\ }\bibfield  {title} {\bibinfo {title} {Chemotaxis in escherichia
  coli analysed by three-dimensional tracking},\ }\href@noop {} {\bibfield
  {journal} {\bibinfo  {journal} {Nature}\ }\textbf {\bibinfo {volume} {239}},\
  \bibinfo {pages} {500} (\bibinfo {year} {1972})}\BibitemShut {NoStop}%
\bibitem [{\citenamefont {De~Gennes}(2004)}]{de2004chemotaxis}%
  \BibitemOpen
  \bibfield  {author} {\bibinfo {author} {\bibfnamefont {P.-G.}\ \bibnamefont
  {De~Gennes}},\ }\bibfield  {title} {\bibinfo {title} {Chemotaxis: the role of
  internal delays},\ }\href@noop {} {\bibfield  {journal} {\bibinfo  {journal}
  {European Biophysics Journal}\ }\textbf {\bibinfo {volume} {33}},\ \bibinfo
  {pages} {691} (\bibinfo {year} {2004})}\BibitemShut {NoStop}%
\bibitem [{\citenamefont {Locsei}(2007)}]{Locsei2007}%
  \BibitemOpen
  \bibfield  {author} {\bibinfo {author} {\bibfnamefont {J.~T.}\ \bibnamefont
  {Locsei}},\ }\bibfield  {title} {\bibinfo {title} {Persistence of direction
  increases the drift velocity of run and tumble chemotaxis},\ }\href
  {https://doi.org/10.1007/s00285-007-0080-z} {\bibfield  {journal} {\bibinfo
  {journal} {Journal of Mathematical Biology}\ }\textbf {\bibinfo {volume}
  {55}},\ \bibinfo {pages} {41–60} (\bibinfo {year} {2007})}\BibitemShut
  {NoStop}%
\bibitem [{\citenamefont {Taktikos}\ \emph {et~al.}(2013)\citenamefont
  {Taktikos}, \citenamefont {Stark},\ and\ \citenamefont
  {Zaburdaev}}]{Taktikos2013}%
  \BibitemOpen
  \bibfield  {author} {\bibinfo {author} {\bibfnamefont {J.}~\bibnamefont
  {Taktikos}}, \bibinfo {author} {\bibfnamefont {H.}~\bibnamefont {Stark}},\
  and\ \bibinfo {author} {\bibfnamefont {V.}~\bibnamefont {Zaburdaev}},\
  }\bibfield  {title} {\bibinfo {title} {How the motility pattern of bacteria
  affects their dispersal and chemotaxis},\ }\href
  {https://doi.org/10.1371/journal.pone.0081936} {\bibfield  {journal}
  {\bibinfo  {journal} {PLoS ONE}\ }\textbf {\bibinfo {volume} {8}},\ \bibinfo
  {pages} {e81936} (\bibinfo {year} {2013})}\BibitemShut {NoStop}%
\bibitem [{\citenamefont {Thornton}\ \emph {et~al.}(2020)\citenamefont
  {Thornton}, \citenamefont {Butler}, \citenamefont {Davis}, \citenamefont
  {Baxter},\ and\ \citenamefont {Wilson}}]{Thornton2020}%
  \BibitemOpen
  \bibfield  {author} {\bibinfo {author} {\bibfnamefont {K.~L.}\ \bibnamefont
  {Thornton}}, \bibinfo {author} {\bibfnamefont {J.~K.}\ \bibnamefont
  {Butler}}, \bibinfo {author} {\bibfnamefont {S.~J.}\ \bibnamefont {Davis}},
  \bibinfo {author} {\bibfnamefont {B.~K.}\ \bibnamefont {Baxter}},\ and\
  \bibinfo {author} {\bibfnamefont {L.~G.}\ \bibnamefont {Wilson}},\ }\bibfield
   {title} {\bibinfo {title} {Haloarchaea swim slowly for optimal chemotactic
  efficiency in low nutrient environments},\ }\bibfield  {journal} {\bibinfo
  {journal} {Nature Communications}\ }\textbf {\bibinfo {volume} {11}},\ \href
  {https://doi.org/10.1038/s41467-020-18253-7} {10.1038/s41467-020-18253-7}
  (\bibinfo {year} {2020})\BibitemShut {NoStop}%
\bibitem [{\citenamefont {Chatterjee}\ \emph {et~al.}(2011)\citenamefont
  {Chatterjee}, \citenamefont {da~Silveira},\ and\ \citenamefont
  {Kafri}}]{chatterjee2011chemotaxis}%
  \BibitemOpen
  \bibfield  {author} {\bibinfo {author} {\bibfnamefont {S.}~\bibnamefont
  {Chatterjee}}, \bibinfo {author} {\bibfnamefont {R.~A.}\ \bibnamefont
  {da~Silveira}},\ and\ \bibinfo {author} {\bibfnamefont {Y.}~\bibnamefont
  {Kafri}},\ }\bibfield  {title} {\bibinfo {title} {Chemotaxis when bacteria
  remember: drift versus diffusion},\ }\href@noop {} {\bibfield  {journal}
  {\bibinfo  {journal} {PLoS computational biology}\ }\textbf {\bibinfo
  {volume} {7}} (\bibinfo {year} {2011})}\BibitemShut {NoStop}%
\bibitem [{\citenamefont {Ghosh}\ \emph {et~al.}(2014)\citenamefont {Ghosh},
  \citenamefont {H\"{a}nggi}, \citenamefont {Marchesoni},\ and\ \citenamefont
  {Nori}}]{Ghosh2014}%
  \BibitemOpen
  \bibfield  {author} {\bibinfo {author} {\bibfnamefont {P.~K.}\ \bibnamefont
  {Ghosh}}, \bibinfo {author} {\bibfnamefont {P.}~\bibnamefont {H\"{a}nggi}},
  \bibinfo {author} {\bibfnamefont {F.}~\bibnamefont {Marchesoni}},\ and\
  \bibinfo {author} {\bibfnamefont {F.}~\bibnamefont {Nori}},\ }\bibfield
  {title} {\bibinfo {title} {Giant negative mobility of janus particles in a
  corrugated channel},\ }\bibfield  {journal} {\bibinfo  {journal} {Physical
  Review E}\ }\textbf {\bibinfo {volume} {89}},\ \href
  {https://doi.org/10.1103/physreve.89.062115} {10.1103/physreve.89.062115}
  (\bibinfo {year} {2014})\BibitemShut {NoStop}%
\bibitem [{\citenamefont {Rizkallah}\ \emph {et~al.}(2023)\citenamefont
  {Rizkallah}, \citenamefont {Sarracino}, \citenamefont {Bénichou},\ and\
  \citenamefont {Illien}}]{Rizkallah2023}%
  \BibitemOpen
  \bibfield  {author} {\bibinfo {author} {\bibfnamefont {P.}~\bibnamefont
  {Rizkallah}}, \bibinfo {author} {\bibfnamefont {A.}~\bibnamefont
  {Sarracino}}, \bibinfo {author} {\bibfnamefont {O.}~\bibnamefont
  {Bénichou}},\ and\ \bibinfo {author} {\bibfnamefont {P.}~\bibnamefont
  {Illien}},\ }\bibfield  {title} {\bibinfo {title} {Absolute negative mobility
  of an active tracer in a crowded environment},\ }\bibfield  {journal}
  {\bibinfo  {journal} {Physical Review Letters}\ }\textbf {\bibinfo {volume}
  {130}},\ \href {https://doi.org/10.1103/physrevlett.130.218201}
  {10.1103/physrevlett.130.218201} (\bibinfo {year} {2023})\BibitemShut
  {NoStop}%
\bibitem [{\citenamefont {Straube}\ and\ \citenamefont
  {H\"{o}fling}(2023)}]{straube23}%
  \BibitemOpen
  \bibfield  {author} {\bibinfo {author} {\bibfnamefont {A.~V.}\ \bibnamefont
  {Straube}}\ and\ \bibinfo {author} {\bibfnamefont {F.}~\bibnamefont
  {H\"{o}fling}},\ }\href {https://doi.org/10.48550/ARXIV.2306.09150} {\bibinfo
  {title} {Depinning transition of self-propelled particles}} (\bibinfo {year}
  {2023})\BibitemShut {NoStop}%
\bibitem [{\citenamefont {Berg}\ and\ \citenamefont
  {Tedesco}(1975)}]{berg1975transient}%
  \BibitemOpen
  \bibfield  {author} {\bibinfo {author} {\bibfnamefont {H.~C.}\ \bibnamefont
  {Berg}}\ and\ \bibinfo {author} {\bibfnamefont {P.}~\bibnamefont {Tedesco}},\
  }\bibfield  {title} {\bibinfo {title} {Transient response to chemotactic
  stimuli in escherichia coli},\ }\href@noop {} {\bibfield  {journal} {\bibinfo
   {journal} {Proceedings of the National Academy of Sciences}\ }\textbf
  {\bibinfo {volume} {72}},\ \bibinfo {pages} {3235} (\bibinfo {year}
  {1975})}\BibitemShut {NoStop}%
\bibitem [{\citenamefont {Goy}\ \emph {et~al.}(1977)\citenamefont {Goy},
  \citenamefont {Springer},\ and\ \citenamefont {Adler}}]{goy1977sensory}%
  \BibitemOpen
  \bibfield  {author} {\bibinfo {author} {\bibfnamefont {M.~F.}\ \bibnamefont
  {Goy}}, \bibinfo {author} {\bibfnamefont {M.~S.}\ \bibnamefont {Springer}},\
  and\ \bibinfo {author} {\bibfnamefont {J.}~\bibnamefont {Adler}},\ }\bibfield
   {title} {\bibinfo {title} {Sensory transduction in escherichia coli: role of
  a protein methylation reaction in sensory adaptation},\ }\href@noop {}
  {\bibfield  {journal} {\bibinfo  {journal} {Proceedings of the National
  Academy of Sciences}\ }\textbf {\bibinfo {volume} {74}},\ \bibinfo {pages}
  {4964} (\bibinfo {year} {1977})}\BibitemShut {NoStop}%
\bibitem [{\citenamefont {Hansen}\ \emph {et~al.}(2008)\citenamefont {Hansen},
  \citenamefont {Endres},\ and\ \citenamefont
  {Wingreen}}]{hansen2008chemotaxis}%
  \BibitemOpen
  \bibfield  {author} {\bibinfo {author} {\bibfnamefont {C.~H.}\ \bibnamefont
  {Hansen}}, \bibinfo {author} {\bibfnamefont {R.~G.}\ \bibnamefont {Endres}},\
  and\ \bibinfo {author} {\bibfnamefont {N.~S.}\ \bibnamefont {Wingreen}},\
  }\bibfield  {title} {\bibinfo {title} {Chemotaxis in escherichia coli: a
  molecular model for robust precise adaptation},\ }\href@noop {} {\bibfield
  {journal} {\bibinfo  {journal} {PLoS Comput Biol}\ }\textbf {\bibinfo
  {volume} {4}},\ \bibinfo {pages} {e1} (\bibinfo {year} {2008})}\BibitemShut
  {NoStop}%
\bibitem [{\citenamefont {Levin}\ \emph {et~al.}(2002)\citenamefont {Levin},
  \citenamefont {Shimizu},\ and\ \citenamefont {Bray}}]{levin2002binding}%
  \BibitemOpen
  \bibfield  {author} {\bibinfo {author} {\bibfnamefont {M.~D.}\ \bibnamefont
  {Levin}}, \bibinfo {author} {\bibfnamefont {T.~S.}\ \bibnamefont {Shimizu}},\
  and\ \bibinfo {author} {\bibfnamefont {D.}~\bibnamefont {Bray}},\ }\bibfield
  {title} {\bibinfo {title} {Binding and diffusion of cher molecules within a
  cluster of membrane receptors},\ }\href@noop {} {\bibfield  {journal}
  {\bibinfo  {journal} {Biophysical journal}\ }\textbf {\bibinfo {volume}
  {82}},\ \bibinfo {pages} {1809} (\bibinfo {year} {2002})}\BibitemShut
  {NoStop}%
\bibitem [{\citenamefont {Endres}\ and\ \citenamefont
  {Wingreen}(2006)}]{endres2006precise}%
  \BibitemOpen
  \bibfield  {author} {\bibinfo {author} {\bibfnamefont {R.~G.}\ \bibnamefont
  {Endres}}\ and\ \bibinfo {author} {\bibfnamefont {N.~S.}\ \bibnamefont
  {Wingreen}},\ }\bibfield  {title} {\bibinfo {title} {Precise adaptation in
  bacterial chemotaxis through “assistance neighborhoods”},\ }\href@noop {}
  {\bibfield  {journal} {\bibinfo  {journal} {Proceedings of the National
  Academy of Sciences}\ }\textbf {\bibinfo {volume} {103}},\ \bibinfo {pages}
  {13040} (\bibinfo {year} {2006})}\BibitemShut {NoStop}%
\bibitem [{\citenamefont {Li}\ and\ \citenamefont
  {Hazelbauer}(2005)}]{li2005adaptational}%
  \BibitemOpen
  \bibfield  {author} {\bibinfo {author} {\bibfnamefont {M.}~\bibnamefont
  {Li}}\ and\ \bibinfo {author} {\bibfnamefont {G.~L.}\ \bibnamefont
  {Hazelbauer}},\ }\bibfield  {title} {\bibinfo {title} {Adaptational
  assistance in clusters of bacterial chemoreceptors},\ }\href@noop {}
  {\bibfield  {journal} {\bibinfo  {journal} {Molecular microbiology}\ }\textbf
  {\bibinfo {volume} {56}},\ \bibinfo {pages} {1617} (\bibinfo {year}
  {2005})}\BibitemShut {NoStop}%
\bibitem [{\citenamefont {Kim}\ \emph {et~al.}(2002)\citenamefont {Kim},
  \citenamefont {Wang},\ and\ \citenamefont {Kim}}]{kim2002dynamic}%
  \BibitemOpen
  \bibfield  {author} {\bibinfo {author} {\bibfnamefont {S.-H.}\ \bibnamefont
  {Kim}}, \bibinfo {author} {\bibfnamefont {W.}~\bibnamefont {Wang}},\ and\
  \bibinfo {author} {\bibfnamefont {K.~K.}\ \bibnamefont {Kim}},\ }\bibfield
  {title} {\bibinfo {title} {Dynamic and clustering model of bacterial
  chemotaxis receptors: structural basis for signaling and high sensitivity},\
  }\href@noop {} {\bibfield  {journal} {\bibinfo  {journal} {Proceedings of the
  National Academy of Sciences}\ }\textbf {\bibinfo {volume} {99}},\ \bibinfo
  {pages} {11611} (\bibinfo {year} {2002})}\BibitemShut {NoStop}%
\bibitem [{\citenamefont {Feng}\ \emph {et~al.}(1999)\citenamefont {Feng},
  \citenamefont {Lilly},\ and\ \citenamefont {Hazelbauer}}]{feng1999enhanced}%
  \BibitemOpen
  \bibfield  {author} {\bibinfo {author} {\bibfnamefont {X.}~\bibnamefont
  {Feng}}, \bibinfo {author} {\bibfnamefont {A.~A.}\ \bibnamefont {Lilly}},\
  and\ \bibinfo {author} {\bibfnamefont {G.~L.}\ \bibnamefont {Hazelbauer}},\
  }\bibfield  {title} {\bibinfo {title} {Enhanced function conferred on
  low-abundance chemoreceptor trg by a methyltransferase-docking site},\
  }\href@noop {} {\bibfield  {journal} {\bibinfo  {journal} {Journal of
  bacteriology}\ }\textbf {\bibinfo {volume} {181}},\ \bibinfo {pages} {3164}
  (\bibinfo {year} {1999})}\BibitemShut {NoStop}%
\bibitem [{\citenamefont {Wu}\ \emph {et~al.}(1996)\citenamefont {Wu},
  \citenamefont {Li}, \citenamefont {Li}, \citenamefont {Long},\ and\
  \citenamefont {Weis}}]{wu1996receptor}%
  \BibitemOpen
  \bibfield  {author} {\bibinfo {author} {\bibfnamefont {J.}~\bibnamefont
  {Wu}}, \bibinfo {author} {\bibfnamefont {J.}~\bibnamefont {Li}}, \bibinfo
  {author} {\bibfnamefont {G.}~\bibnamefont {Li}}, \bibinfo {author}
  {\bibfnamefont {D.~G.}\ \bibnamefont {Long}},\ and\ \bibinfo {author}
  {\bibfnamefont {R.~M.}\ \bibnamefont {Weis}},\ }\bibfield  {title} {\bibinfo
  {title} {The receptor binding site for the methyltransferase of bacterial
  chemotaxis is distinct from the sites of methylation},\ }\href@noop {}
  {\bibfield  {journal} {\bibinfo  {journal} {Biochemistry}\ }\textbf {\bibinfo
  {volume} {35}},\ \bibinfo {pages} {4984} (\bibinfo {year}
  {1996})}\BibitemShut {NoStop}%
\bibitem [{\citenamefont {Schulmeister}\ \emph {et~al.}(2008)\citenamefont
  {Schulmeister}, \citenamefont {Ruttorf}, \citenamefont {Thiem}, \citenamefont
  {Kentner}, \citenamefont {Lebiedz},\ and\ \citenamefont
  {Sourjik}}]{schulmeister2008protein}%
  \BibitemOpen
  \bibfield  {author} {\bibinfo {author} {\bibfnamefont {S.}~\bibnamefont
  {Schulmeister}}, \bibinfo {author} {\bibfnamefont {M.}~\bibnamefont
  {Ruttorf}}, \bibinfo {author} {\bibfnamefont {S.}~\bibnamefont {Thiem}},
  \bibinfo {author} {\bibfnamefont {D.}~\bibnamefont {Kentner}}, \bibinfo
  {author} {\bibfnamefont {D.}~\bibnamefont {Lebiedz}},\ and\ \bibinfo {author}
  {\bibfnamefont {V.}~\bibnamefont {Sourjik}},\ }\bibfield  {title} {\bibinfo
  {title} {Protein exchange dynamics at chemoreceptor clusters in escherichia
  coli},\ }\href@noop {} {\bibfield  {journal} {\bibinfo  {journal}
  {Proceedings of the National Academy of Sciences}\ }\textbf {\bibinfo
  {volume} {105}},\ \bibinfo {pages} {6403} (\bibinfo {year}
  {2008})}\BibitemShut {NoStop}%
\bibitem [{\citenamefont {Li}\ and\ \citenamefont
  {Hazelbauer}(2004)}]{li2004cellular}%
  \BibitemOpen
  \bibfield  {author} {\bibinfo {author} {\bibfnamefont {M.}~\bibnamefont
  {Li}}\ and\ \bibinfo {author} {\bibfnamefont {G.~L.}\ \bibnamefont
  {Hazelbauer}},\ }\bibfield  {title} {\bibinfo {title} {Cellular stoichiometry
  of the components of the chemotaxis signaling complex},\ }\href@noop {}
  {\bibfield  {journal} {\bibinfo  {journal} {Journal of bacteriology}\
  }\textbf {\bibinfo {volume} {186}},\ \bibinfo {pages} {3687} (\bibinfo {year}
  {2004})}\BibitemShut {NoStop}%
\bibitem [{\citenamefont {Sneddon}\ \emph {et~al.}(2011)\citenamefont
  {Sneddon}, \citenamefont {Faeder},\ and\ \citenamefont
  {Emonet}}]{sneddon2011efficient}%
  \BibitemOpen
  \bibfield  {author} {\bibinfo {author} {\bibfnamefont {M.~W.}\ \bibnamefont
  {Sneddon}}, \bibinfo {author} {\bibfnamefont {J.~R.}\ \bibnamefont
  {Faeder}},\ and\ \bibinfo {author} {\bibfnamefont {T.}~\bibnamefont
  {Emonet}},\ }\bibfield  {title} {\bibinfo {title} {Efficient modeling,
  simulation and coarse-graining of biological complexity with nfsim},\
  }\href@noop {} {\bibfield  {journal} {\bibinfo  {journal} {Nature methods}\
  }\textbf {\bibinfo {volume} {8}},\ \bibinfo {pages} {177} (\bibinfo {year}
  {2011})}\BibitemShut {NoStop}%
\bibitem [{\citenamefont {Stewart}\ \emph {et~al.}(2000)\citenamefont
  {Stewart}, \citenamefont {Jahreis},\ and\ \citenamefont
  {Parkinson}}]{stewart2000rapid}%
  \BibitemOpen
  \bibfield  {author} {\bibinfo {author} {\bibfnamefont {R.~C.}\ \bibnamefont
  {Stewart}}, \bibinfo {author} {\bibfnamefont {K.}~\bibnamefont {Jahreis}},\
  and\ \bibinfo {author} {\bibfnamefont {J.~S.}\ \bibnamefont {Parkinson}},\
  }\bibfield  {title} {\bibinfo {title} {Rapid phosphotransfer to chey from a
  chea protein lacking the chey-binding domain},\ }\href@noop {} {\bibfield
  {journal} {\bibinfo  {journal} {Biochemistry}\ }\textbf {\bibinfo {volume}
  {39}},\ \bibinfo {pages} {13157} (\bibinfo {year} {2000})}\BibitemShut
  {NoStop}%
\bibitem [{\citenamefont {{van Kampen}}(2007)}]{VANKAMPEN2007396}%
  \BibitemOpen
  \bibfield  {author} {\bibinfo {author} {\bibfnamefont {N.}~\bibnamefont {{van
  Kampen}}},\ }\bibfield  {title} {\bibinfo {title} {Chapter xvi - stochastic
  differential equations},\ }in\ \href
  {https://doi.org/https://doi.org/10.1016/B978-044452965-7/50019-2} {\emph
  {\bibinfo {booktitle} {Stochastic Processes in Physics and Chemistry (Third
  Edition)}}},\ \bibinfo {series and number} {North-Holland Personal Library},\
  \bibinfo {editor} {edited by\ \bibinfo {editor} {\bibfnamefont
  {N.}~\bibnamefont {{van Kampen}}}}\ (\bibinfo  {publisher} {Elsevier},\
  \bibinfo {address} {Amsterdam},\ \bibinfo {year} {2007})\ \bibinfo {edition}
  {third edition}\ ed.,\ pp.\ \bibinfo {pages} {396--421}\BibitemShut {NoStop}%
\end{thebibliography}

\end{document}